\documentclass{article}

\usepackage{fullpage}
\usepackage{graphicx}
\graphicspath{{.}}

\usepackage{hyperref}
\usepackage{amssymb}
\usepackage{amsmath}

\usepackage{tikz} 
\usetikzlibrary{backgrounds}
\usetikzlibrary{calc}

\usepackage{natbib}
\bibpunct{(}{)}{;}{a}{}{,}

\definecolor{grey}{rgb}{0.5,0.5,0.5}
\definecolor{lightgrey}{rgb}{0.7,0.7,0.7}
\definecolor{verylightgrey}{rgb}{0.9,0.9,0.9}
\definecolor{darkgrey}{rgb}{0.2,0.2,0.2}

\definecolor{grey5}{rgb}{0.9,0.9,0.9}
\definecolor{grey4}{rgb}{0.833,0.833,0.833}
\definecolor{grey3}{rgb}{0.766,0.766,0.766}
\definecolor{grey2}{rgb}{0.7,0.7,0.7}
\definecolor{grey1}{rgb}{0.533,0.533,0.533}

\title{Relational hyperevent models for the coevolution of coauthoring and citation networks}

 \author{J\"urgen Lerner\\
   University of Konstanz, Germany\\
   \texttt{juergen.lerner@uni-konstanz.de}
         \and
         Marian-Gabriel H\^{a}ncean\\
         University of Bucharest, Romania\\
         \texttt{gabriel.hancean@sas.unibuc.ro}
         \and
         Alessandro Lomi\\
         Universit\`a della Svizzera italiana, Switzerland\\
         \texttt{alessandro.lomi@usi.ch}
 }

 \date{}

\begin{document}

\maketitle

\begin{abstract}
  The development of suitable statistical models for the analysis of bibliographic networks has trailed behind the empirical ambitions expressed by recent studies of science of science. Extant research typically restricts the analytical focus to either paper citation networks, or author collaboration networks. These networks involve not only direct relationships between papers or authors, but also a broader system of dependencies between the references of papers connected through multiple simultaneous citation links. In this work, we extend recently developed relational hyperevent models (RHEM) to analyze scientific networks -- systems of scientific publications connected by citations and authorship. We introduce new covariates that represent theoretically relevant and empirically meaningful sub-network configurations. The new model specification supports testing of hypotheses that align with the polyadic nature of scientific publication events and the multiple interdependencies between authors and references of current and prior papers.  We implement the model using open-source software to analyze a large, publicly available scientific network dataset. A significant finding of the study is the tendency for subsets of papers to be repeatedly cited together across publications. This result is crucial as it suggests that the papers' impact may be partly due to endogenous network processes. More broadly, the study shows that models accounting for both the hyperedge structure of publication events and the interconnections between authors and references significantly enhance our understanding of the network mechanisms that drive scientific production, productivity, and impact.\medskip

\end{abstract}

\section{Introduction}
\label{sec:intro}
Publicly available data on very large networks of scientific citations are reaching unprecedented levels of scale, detail, and scope \citep{lin2023sciscinet}. Parallel progress has been made on the analytical front where statistical models are now available to analyze increasingly large citation networks \citep{filippi2024stochastic}. Stimulated by these developments, citation networks are now broadly recognized as the prime empirical basis for understanding the implications of the network structure and dynamics of scientific production, innovation, and impact \citep{barabasi2002evolution,  meng2023hidden, newman2001structure, radicchi2011citation, mukherjee2017nearly}. 

As \citet[p.~1046]{liu2023data} recently summarized: ``The advent of large-scale data sets that trace the workings of science has encouraged researchers from many different disciplinary backgrounds to turn scientific methods into science itself.'' The progressive development and expansion of the new interdisciplinary field of ``science of science'' has considerably increased the interest in the network analysis of scientific networks -- systems of scientific publications connected by bibliographic citations \citep{fortunato2018science}. 

Studies of science production, productivity and impact based on the analysis of scientific networks are becoming increasingly common, particularly in top interdisciplinary science journals \citep{uzzi2013atypical}. For example, \citet{guimera2005team} explain performance of scientists working together in terms of team assembly mechanisms -- or rules of partner selection. \citet{uzzi2013atypical} analyze how the atypicality of a paper's references explains the probability to become a ``hit paper.'' \citet{park2023papers} found that papers are becoming less disruptive over time by showing a decline in the ``citation disruption index'' (CDI) -- a quantitative measure of the extent to which papers citing a focal paper do not cite its references. These results have been made possible by careful empirical analysis of scientific networks. While different, these works all try to understand the relation between scientific production and ultimate success of scientific ideas in terms of positions that scientific papers occupy in networks of paper citations \citep{sarigol2014predicting}. 

It is generally acknowledged that quantitative measures derived from scientific networks need to control for empirically observed features of these networks \citep{ newman2001structure}. For instance, \citet{uzzi2013atypical} normalize raw co-citation counts by expected values in randomized versions of the citation network, maintaining outdegree (i.\,e., the length of reference lists) and indegree (i.\,e., the number of citations that a paper receives). \citet{park2023papers} adopt similar normalizations. However, randomization techniques for citation networks controlling for outdegrees and indegrees of papers are typically insufficient to provide \emph{general} models for scientific networks and cannot test multiple competing hypotheses explaining scientific collaboration and impact. For instance, \citet{petersen2023disruption} claim that the CDI is confounded by changing citation behavior -- caused, for example, by the increased availability of search engines for scientific publications, which alone can explain the increased tendency for ``triadic closure'' in the citation network.

More generally, scientific networks emerging from patterns of paper citations involve two disjoint classes of entities that may be usefully represented as network nodes:  scientific papers and their authors \citep{carusi2019scientific}.  This implies the need to represent at least two network sub-systems \citep{saavedra2009simple, tumminello2011statistically}: authors connected to the papers they write, and papers connected to previous papers they cite (''backward citations'').  In this view, citation networks involve a number of dual dependencies \citep{breiger1974duality, lee2018doorway}, but extant research has typically restricted attention only to a subset of these dependence relations.  For example, \citet{ferligoj2015scientific} model coauthoring networks with stochastic actor-oriented models (SAOM) \citep{s-mlnd-05}. \citet{fritz2023modelling} adopt temporal exponential random graph models (TERGM) \citep{lkr-ergmsn-13,krivitsky2014separable} to model collaboration among inventors that jointly file patents. Very much in the spirit of the current paper, \citet{espinosa2024co} specify and estimate multilevel SAOM  to examine relations between papers, authors, and research institutions within a scientific community of astronomers from 2013 to 2015.

We start the present study by emphasizing what we see as three constitutive features of scientific networks emerging from time-ordered sequences of individual publication events. First, interaction among nodes (authors and papers) involves multiple parallel counting processes such as, for instance, the number of coauthored papers, or the number of citations to a paper. This property makes point-process models for network data \citep{perry2013point}, often denoted as relational event models (REM) \citep{bianchi2023relational, butts2008relational}, a natural choice for modeling this kind of networks \citep{butts2023relational}. Second, a single publication event does not connect only two nodes at a time, but it connects a varying and potentially large number of authors and cited papers simultaneously, which involves ``polyadic'' (one-to-many) interaction, or ``hyperevents'' \citep{lerner2023micro,lerner2023relational}. Treating these polyadic interactions as collections of dyadic interactions artificially inflates the number of observations, introduces dependence among observations by construction, and deprives analysts of the possibility to test for theoretically relevant and empirically plausible ``higher-order effects'' \citep{lerner2023relational}. Third, publication events connect nodes of different types (authors and papers) with relations of different types: authors are connected to the papers they write (alone or in teams), and papers are connected to the multiple papers they cite. We argue that modeling either of these relations separately affords only a partial understanding of the dynamics of scientific networks.

Our main objective in this paper is to derive and test appropriate relational event models that incorporate these core features of scientific networks and that support hypotheses testing about a variety of empirically plausible and theoretically meaningful data generating mechanisms. We narrow the focus of our attention on relational event models because, unlike better established exponential random graph models (ERGM) \citep{amati2018social}, temporal ERGMS \citep{desmarais2012statistical} or stochastic actor-oriented models (SAOMS) \citep{snijders2017stochastic}, relational event models (REM) can deal with fine-grained time information thus removing the need to aggregate observations over arbitrary time intervals \citep{bianchi2023relational}.  

We contribute to  -- but also go beyond extant research in at least two ways.  First, while current studies have recognized the analytical value of treating scientific networks as sequences of time-ordered sequences of relational hyperevents, we still know little about the underlying relational mechanisms that may be responsible for the observed structure of scientific networks. For example, \citet{lerner2023micro}, model coauthoring networks via RHEM in an attempt to  explain the ``impact'' of papers, where  impact is reconstructed in terms of a single numerical variable:  the cumulative number of citations received at the time of data collection. However, the process through which papers accumulate citations was not modeled. In consequence, the model proposed by \citet{lerner2023micro} cannot be adopted to test or control for competing explanations of reference selection. In this paper we take a more constructive approach by modeling directly the generating process underlying the observable network dynamics of paper citations. 

Second, relational hyperevent models are currently available for the analysis of processes defined on networks with a single class of nodes (one-mode, or unipartite networks) \citep{lerner2023relational}, or with two classes of nodes, e.\,g., participants and ``labels'' of the event, where interaction is only between the node classes (two-mode, or bipartite networks) \citep{bright2024offence} -- but not both simultaneously. In consequence, available RHEM cannot be specified to analyze mixed-mode networks \citep{snijders2013model}, where relational events can occur -- and may be observed both between nodes in different classes, as well as nodes within of the same class. The value of such a model should be obvious in the analysis of scientific networks -- the empirical setting of our study -- where scientific collaboration produce direct relations between multiple authors, citations produce relations between multiple papers and their references, and publications produce associations between papers and their authors -- and hence their institutional affiliation \citep{espinosa2024co}. In this paper we extend the capacity of RHEM to analyze dynamic mixed-mode network processes.  The corresponding effects that we specify and test are not only theoretically interesting, but also empirically important --as the results we report clearly demonstrate.

The empirical value of the model is tested using a large publicly available data on a scientific network containing more than a million papers and authors. We use this sample to test for the presence of higher-order dependencies and mixed two-mode dependencies and to compare the relative strength of the various effects, measured in terms of their individual contribution to the log-likelihood. 

We show that the magnitude of the new effects that we specify to capture salient aspects of mixed-mode network processes is comparable to some of the strongest effects documented by extant studies of paper citations based on one-mode network data.  The exclusive modeling of citation networks -- disregarding the interdependent coauthoring network -- ignores that the choice of references cited in published papers is highly dependent on the authors of those papers. Plausible patterns in the coevolution of coauthoring and citation networks include the tendency of authors to: (i) cite their own prior papers \citep{seeber2019self}; (ii) cite papers of their past coauthors \citep{martin2013coauthorship}; (iii) adopt citations of their past coauthors \citep{goldberg2015modelling}; (iv) repeatedly cite the same papers \citep{price1976general, merton1968matthew}, and (v) repeatedly cite papers written by the same authors \citep{white2001authors}. Our analysis reveals a significant propensity of subsets of references to self-reproduce across publications. This result supports the conjecture that references come, at least to some extent, in discrete ``packages'' of variable size. To the extent that  the ``impact'' of a scientific publication depends on the number of citations received \citep{radicchi2008universality}, this results is important because it suggests that individual impact may be driven, in part, by non-individual components.  

We conclude our study inviting readers to reflect on the potential far-reaching implications, both pragmatic as well as conceptual, of the results afforded by the new models we present. These models can now be specified and estimated on very large data sets produced by scientific networks and connected systems such as, for example, patent citation systems \citep{filippi2024stochastic}. 

The results of the analysis that we report were obtained using a publicly available, open-source software\footnote{\url{https://github.com/juergenlerner/eventnet}} that interested readers can access freely. The combination of publicly available data and freely available software contributes to making the results of the analysis we present in this paper fully reproducible.

\section{Models}
\label{sec:models}

The models considered in this paper are defined within the point-process framework for relational event data proposed by \citet{perry2013point}. Theorems about the consistency and asymptotic normality of the maximum partial likelihood estimators (MPLE) presented in \citet{perry2013point} also apply to the likelihood functions discussed below.  Building  directly on prior work \citep{perry2013point,lerner2023micro,lerner2023relational}, we want to extend models and develop novel hyperedge covariates, capturing the polyadic structure of coauthoring and citation and, at the same time, the interrelation between these two kinds of networks.

We propose models for counting processes $N_{(I,J)}(t)$, where $I$ is a \emph{set} of authors, publishing papers that cite a \emph{set} of references $J$. Thereby, we extend existing REM for citation networks \citep{hunter2011dynamic} in two directions. First, we do not focus exclusively on the cited papers, but consider also the interrelation between authors in $I$ and references in $J$. Secondly, we take into account that coauthoring and citation networks cannot be completely represented in terms of purely dyadic interactions, connecting only two nodes at a time. Note that a single publication event entails a varying number of authors citing a varying -- and arbitrarily large -- number of references.

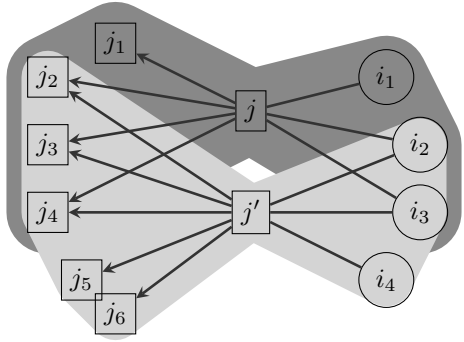
\begin{figure}
  \begin{center}
    \begin{tikzpicture}[scale=0.9,>=stealth]
      \tikzstyle{actor}=[circle,minimum size=1mm,draw=black];
      \tikzstyle{work}=[rectangle,minimum size=1mm,draw=black];
      \node[work] (w) at (0,0) {$j$};
      \node[work] (w1) at (-2.0,1.0) {$j_1$};
      \node[work] (w2) at (-3.0,0.5) {$j_2$};
      \node[work] (w3) at (-3.0,-0.5) {$j_3$};
      \node[work] (w4) at (-3.0,-1.5) {$j_4$};
      \node[work] (w5) at (-2.5,-2.5) {$j_5$};
      \node[work] (w6) at (-2.0,-3.0) {$j_6$};
      \node[work] (wt2) at (0,-1.5) {$j'$};
      \node[actor] (a1) at (2,0.5) {$i_1$};
      \node[actor] (a2) at (2.5,-0.5) {$i_2$};
      \node[actor] (a3) at (2.5,-1.5) {$i_3$};
      \node[actor] (a4) at (2.0,-2.5) {$i_4$};
      
      \draw[color=darkgrey,line width=1pt] (w) to (a1);
      \draw[color=darkgrey,line width=1pt] (w) to (a2);
      \draw[color=darkgrey,line width=1pt] (w) to (a3);
      \draw[color=darkgrey,line width=1pt] (wt2) to (a2);
      \draw[color=darkgrey,line width=1pt] (wt2) to (a3);
      \draw[color=darkgrey,line width=1pt] (wt2) to (a4);

      \draw[color=darkgrey,line width=1pt,->] (w) to (w1);
      \draw[color=darkgrey,line width=1pt,->] (w) to (w2);
      \draw[color=darkgrey,line width=1pt,->] (w) to (w3);
      \draw[color=darkgrey,line width=1pt,->] (w) to (w4);

      \draw[color=darkgrey,line width=1pt,->] (wt2) to (w2);
      \draw[color=darkgrey,line width=1pt,->] (wt2) to (w3);
      \draw[color=darkgrey,line width=1pt,->] (wt2) to (w4);
      \draw[color=darkgrey,line width=1pt,->] (wt2) to (w5);
      \draw[color=darkgrey,line width=1pt,->] (wt2) to (w6);

      \begin{pgfonlayer}{background}
        \foreach \nodename in {w,w1,w2,w3,w4,w5,w6,wt2,a1,a2,a3,a4} {
          \coordinate (\nodename') at (\nodename);
        }
        \path[fill=grey1,draw=grey1,line width=1.1cm, line cap=round, line join=round] 
        (w') to (w1') to (w2') to (w3') to (w4')  to (w') to (a1') to (a2') to (a3') to (w') -- cycle;
        \path[fill=grey4,draw=grey4,line width=0.7cm, line cap=round, line join=round] 
        (wt2') to (w2') to (w3') to (w4') to (w5') to (w6')  to (wt2') to (a4') to (a3') to (a2') to (wt2') -- cycle;
      \end{pgfonlayer}
    \end{tikzpicture}
  \end{center}
  \caption{\label{fig:event} Publication event of paper $j$, authored by $i_1,\dots,i_3$ and citing papers $j_1,\dots,j_4$, followed by the publication event of paper $j'$, authored by $i_2,\dots,i_4$ and citing papers $j_2,\dots,j_6$. The two authors $i_2,i_3$ repeatedly cite the three papers $j_2,j_3,j_4$. Mixed two-mode hyperedges, containing the published paper, its authors, and its references, are displayed as gray-shaded areas enclosing the participating nodes.}
\end{figure}

Let $\mathcal{I}$ be a finite set of authors and $\mathcal{J}$ be a finite set of papers. For a point in time $t>0$, let $\mathcal{I}_t\subseteq\mathcal{I}$ be those authors who could potentially publish a paper at $t$ and let $\mathcal{J}_t\subseteq\mathcal{J}$ be the set of those papers that could potentially be cited by a paper published at $t$ (``prior work''). Models defined below specify distributions for sequences of publication events $(t_1,j_1,I_1,J_1),\dots,(t_n,j_n,I_n,J_n)$, where $(t_m,j_m,I_m,J_m)$ indicates that authors $I_m\subseteq\mathcal{I}_{t_m}$ publish paper $j_m\in\mathcal{J}$ at time $t_m$, citing papers $J_m\subseteq\mathcal{J}_{t_m}$ in their references. See Fig.~\ref{fig:event} for the graphical illustration of a publication event, followed by a later publication whose reference list could partially be explained by the tendency of authors to repeatedly cite the same papers. 

For any set of authors $I\subseteq\mathcal{I}_t$ (that is, a possible set of coauthors) and any set of papers $J\subseteq\mathcal{J}_t$ (that is, a possible set of references), let $x_t(I,J)$ be a $p$-dimensional vector of covariates with associated unknown vector of parameters $\beta_0$.

\subsection{The joint model}
\label{sec:joint_model}

Consider the counting processes on $\mathbb{R}_+\times\mathcal{P}(\mathcal{I})\times\mathcal{P}(\mathcal{J})$, where the intensity on $(I,J)\in\mathcal{P}(\mathcal{I})\times\mathcal{P}(\mathcal{J})$ is specified as
\begin{equation}
  \label{eq:lambda}
\lambda_t(I,J)=\overline{\lambda}_t(|I|,|J|)\exp\left\{\beta_0^{\rm T}x_t(I,J)\right\}{\bf 1}\{I\subseteq\mathcal{I}_t\wedge J\subseteq\mathcal{J}_t\}\enspace.
\end{equation}
For a pair of positive integers $K$ and $L$, giving the number of authors and the number of references, respectively, $\overline{\lambda}_t(K,L)$ is the baseline intensity, and ${\bf 1}\{I\subseteq\mathcal{I}_t\wedge J\subseteq\mathcal{J}_t\}$ is the indicator whether the pair $(I,J)$ is ``at risk'' of experiencing an event at time $t$. For a pair $(I,J)$ that is at risk, the intensity $\lambda_t(I,J)$ is the product of the baseline intensity with the relative rate, or hazard ratio, $\exp\left\{\beta_0^{\rm T}x_t(I,J)\right\}$, whose estimation is of central interest for empirical applications. The hazard ratio indicates  the factor by which a possible publication event at time $t$ with authors $I$ and references $J$ is more or less likely than an alternative publication event with authors $I'$ and references $J'$. The hyperedge covariates $x_t(I,J)$ can be specified as functions of exogenous attributes of authors or papers and/or they can be functions of the history of the process, that is, of publication events at time $t'<t$.  In the next section (``Effects'') 
we offer a representative list of possible hyperedge covariates.

Note that the specification in (\ref{eq:lambda}) includes the number of authors $|I|$ and the number of references $|J|$ in the baseline rate, implying that the hazard ratio reveals the factor by which observed events $(I,J)$ are more or less likely than alternative events $(I',J')$ with the same number of authors ($|I'|=|I|$) and the same number of references ($|J'|=|J|$). This decision is in line with -- but extends -- the specification in \citet{perry2013point,lerner2023relational} who include the number of receivers in the baseline rate (the number of senders is fixed to one in these papers). Ultimately, this decision can be motivated by the simple but crucial observation that the number of subsets of size $k$ increases exponentially in $k$, as long as $k$ is much smaller than the size of the entire set (i.\,e., the number of all authors who could publish or the number of all citable papers, respectitively). For instance, the baseline publication rate on sets comprising ten authors is tremendously lower than the baseline rate on sets comprising two authors -- simply because there are so many more different author sets of size ten, than sets of size two. The decision to absorb the number of authors and references in the baseline rate ensures that effects of substantive interest, shaping the hazard ratio, do not get confounded by trivial but very strong variation in the number of subsets of given sizes.

The specification given in (\ref{eq:lambda}) is a stratified Cox proportional hazard model \citep{cox1972regression,aalen2008survival}, where the strata are defined by the combination of the number of authors and the number of references. This simple observation is very useful since it implies that model parameters can be estimated by standard software for Cox regression, as it will be explained in more detail in the empirical section of this paper.

Let $(t_1,j_1,I_1,J_1),\dots,(t_n,j_n,I_n,J_n)$ be the observed sequence of publication events. Adapting models proposed in \citet{perry2013point,lerner2023relational} to the case of publication events, the model defined in (\ref{eq:lambda}) leads to the following partial likelihood at time $t$, evaluated at parameters $\beta$. 
\begin{equation}
  \label{eq:likelihood}
  L_t(\beta)=\prod_{t_m\leq t}\frac{
  \exp\left\{\beta^{\rm T}x_{t_m}(I_m,J_m)\right\}}
  {\sum_{(I,J)\in{\mathcal{I}_{t_m}\choose |I_m|}\times{\mathcal{J}_{t_m}\choose |J_m|}}
    \exp\left\{\beta^{\rm T}x_{t_m}(I,J)\right\}
  }
\end{equation}
(For any set $M$ and integer $k$, we denote by $|M|$ the cardinality of $M$ and we denote by $\binom{M}{k}=\{M'\subseteq M\,;\;|M'|=k\}$ the set of all subsets of size $k$ of $M$.)
Theorems from \citet{perry2013point} imply that, under generally accepted assumptions, estimating parameters $\hat{\beta}$ by maximizing the partial likelihood (\ref{eq:likelihood}) produces consistent estimates.\footnote{To see that this claim holds, consider the comment by \citet[][p.~828]{perry2013point}: ``\emph{the multicast case can be considered as a special case of the single-receiver case}.'' In our case, events have not only several ``receivers'' (cited papers) but also several ``senders'' (authors), which does not affect the applicability of their theorems.}

Equation (\ref{eq:likelihood}) is the partial likelihood of a stratified CoxPH model (see above) and standard maximum likelihood estimation of the parameters $\beta$ relies on the assumption that events are conditionally independent, given the past events, and that the dependence on the past is completely captured by the vector of covariates $x_t$. We note that the validity of this assumption in any given empirical data is, in general, unprovable and it is unlikely to hold in the strict sense in complex, large-scale data, as the one  analyzed in this paper. \citet{aalen2008survival} note that if not all dependence on the past is captured by the given covariates, then: ``\emph{Often the estimation procedures are still valid, but the variance estimates from martingale theory are not correct (they will typically underestimate the true variance) and have to be substituted with sandwich type estimators.}'' \citep[p.~303]{aalen2008survival} Following this recommendation, we will estimate parameters with robust estimation in the empirical section of this paper.

For a large sample of authors and papers, the number of subsets in ${\mathcal{I}_{t_m}\choose |I_m|}$ and ${\mathcal{J}_{t_m}\choose |J_m|}$ grows exponentially in $|I_m|$, or $|J_m|$, respectively, leading to a computationally intractable likelihood for all but the smallest numbers of coauthors and references. We therefore resort to the approach of nested case-control sampling for CoxPH models \citep{bgl-mascdcphm-95,keogh2014nested}, which has also been applied for the practical estimation of large REM and RHEM, for instance, in \citet{lerner2020reliability,lerner2023relational}. For a positive integer $q$, representing the number of non-events (or ``controls'') per event (or ``case''), let $\tilde{\mathcal{R}}_t(\mathcal{I}_t,I,\mathcal{J}_t,J,q)\subseteq{\mathcal{I}_{t}\choose |I|}\times{\mathcal{J}_{t}\choose |J|}$ be a set of pairs of subsets that is sampled uniformly at random from
\begin{equation}
  \label{eq:sampled_risk_set}
  \left\{
  \mathcal{R}\subseteq
          {\mathcal{I}_{t}\choose |I|}\times{\mathcal{J}_{t}\choose |J|}
          \,;\;(I,J)\in\mathcal{R}\wedge|\mathcal{R}|=q+1
  \right\}
\end{equation}
The sampled risk set $\tilde{\mathcal{R}}_t(\mathcal{I}_t,I,\mathcal{J}_t,J,q)$ can be constructed by including $(I,J)$ (the observed pair of authors and references, that is, the ``case'') plus $q$ alternative pairs $(I',J')\neq (I,J)$, the ``controls'', sampled without replacement uniformly and independently at random from ${\mathcal{I}_{t}\choose |I|}\times{\mathcal{J}_{t}\choose |J|}$.

We obtain a sampled partial likelihood $\tilde{L}_t(\beta)$ by substituting the sampled risk set $\tilde{\mathcal{R}}_{t_m}(\mathcal{I}_{t_m},I_m,\mathcal{J}_{t_m},J_m,q)$ for ${\mathcal{I}_{t_m}\choose |I_m|}\times{\mathcal{J}_{t_m}\choose |J_m|}$ in the summation of (\ref{eq:likelihood}):
\begin{equation}
  \label{eq:likelihood_sampled}
  \tilde{L}_t(\beta)=\prod_{t_m\leq t}\frac{
    \exp\left\{\beta^{\rm T}x_{t_m}(I_m,J_m)\right\}
    }
    {\sum_{(I,J)\in\tilde{\mathcal{R}}_{t_m}(\mathcal{I}_{t_m},I_m,\mathcal{J}_{t_m},J_m,q)}
    \exp\left\{\beta^{\rm T}x_{t_m}(I,J)\right\}
    }
\end{equation}
According  to \citet{bgl-mascdcphm-95}, estimating parameters $\hat{\beta}$ by maximizing $\tilde{L}_t(\beta)$ is a consistent estimator under reasonable conditions (also compare \citet{lerner2023relational}). As noted above, parameter estimation can be done by standard software for Cox regression -- provided that the software is given a table containing the values of the hyperedge covariates $x_t(I,J)$ for all instances, that is, for all events and sampled controls. We provide details on this aspect in the empirical section of this paper.

\subsection{Separation into coauthoring model and citation model}
\label{sec:separate_models}

For some empirical applications it may be useful to decompose the joint model into two sub-models: a ``coauthoring model'' explaining the propensity of sets of authors to coauthor papers and a ``citation model'' explaining the conditional probability of the reference list of a paper published by a given set of authors. In some applications this decomposition may more faithfully reflect scientific collaboration and publication processes in which scientists first form teams and then write their common paper (which includes the decision about which papers to cite). The decomposition of the joint model into the two submodels reflects this separation and could potentially support a simpler interpretation of the different effects in the model. This decomposition is a possible option in modeling scientific networks, although it is not strictly necessary. We emphasize that this model decomposition, if adopted, would still maintain the interaction of coauthoring and citation networks. For instance, coauthoring probabilities can still be modeled dependent on past citations and the conditional probability of reference lists can still be modeled dependent on the authors of the current and past publications.

Specifications of the coauthoring model and citation model are very similar to the joint model -- except that the space of possibilities in these models are sets of authors (for the coauthoring model), or sets of papers (for the citation model), respectively. We provide details in the appendix.

\section{Effects}
\label{sec:effects}

Effects in RHEM for scientific citation networks are operationalized by specifying hyperedge covariates $x_t(I,J)$. These covariates can be functions of exogenous attributes of nodes (authors or papers) or they can be functions of the history of the process, that is, functions of publication events $(t',j',I',J')$ with $t'<t$. These history-dependent covariates are more demanding to understand, but also introduce empirically more interesting effects in the models. We will therefore focus on the definition of history-dependent covariates, which will proceed in two steps. First, we define a small number of time-dependent functions (denoted as ``network attributes'') assigning values to nodes or sets of nodes, summarizing past interaction in prior events. These network attributes represent the state of the scientific network at any point in time. Second, we define the model covariates as functions of these network attributes. While this separation is not necessary, we argue that the network attributes are conceptually intuitive summaries of the interaction history, reduce the cognitive burden to understand the definition of covariates (in other words, they reduce the complexity of formulas), and represent a modular approach in which definitions of attributes might be modified without changing the definition of covariates as functions of attributes, and vice versa.

\begin{figure}
  \begin{center}
    \begin{tikzpicture}
      \node (image) at (-5,0) {%
        \begin{tikzpicture}[scale=0.9,>=stealth]
          \tikzstyle{actor}=[circle,minimum size=1mm,draw=black];
          \tikzstyle{work}=[rectangle,minimum size=1mm,draw=black];
          \node[work] (w) at (0,0) {$j$};
          \node[work] (w1) at (-2.0,1.0) {$j_1$};
          \node[work] (w2) at (-3.0,0.5) {$j_2$};
          \node[work] (w3) at (-3.0,-0.5) {$j_3$};
          \node[work] (w4) at (-3.0,-1.5) {$j_4$};
          \node[work] (w5) at (-2.5,-2.5) {$j_5$};
          \node[work] (w6) at (-2.0,-3.0) {$j_6$};
          \node[work] (wt2) at (0,-1.5) {$j'$};
          \node[actor] (a1) at (2,0.5) {$i_1$};
          \node[actor] (a2) at (2.5,-0.5) {$i_2$};
          \node[actor] (a3) at (2.5,-1.5) {$i_3$};
          \node[actor] (a4) at (2.0,-2.5) {$i_4$};
          \node[actor] (a5) at (-4.5,-2.0) {$i_5$};
          \node[actor] (a6) at (-4.0,-3.0) {$i_6$};
          
          \draw[color=darkgrey,line width=1pt] (w) to (a1);
          \draw[color=darkgrey,line width=1pt] (w) to (a2);
          \draw[color=darkgrey,line width=1pt] (w) to (a3);
          \draw[color=darkgrey,line width=1pt] (wt2) to (a2);
          \draw[color=darkgrey,line width=1pt] (wt2) to (a3);
          \draw[color=darkgrey,line width=1pt] (wt2) to (a4);

          \draw[color=darkgrey,line width=1pt] (w4) to (a5);
          \draw[color=darkgrey,line width=1pt] (w5) to (a5);
          \draw[color=darkgrey,line width=1pt] (w5) to (a6);
          \draw[color=darkgrey,line width=1pt] (w6) to (a6);
          
          \draw[color=darkgrey,line width=1pt,->] (w) to (w1);
          \draw[color=darkgrey,line width=1pt,->] (w) to (w2);
          \draw[color=darkgrey,line width=1pt,->] (w) to (w3);
          \draw[color=darkgrey,line width=1pt,->] (w) to (w4);

          \draw[color=darkgrey,line width=1pt,->] (wt2) to (w2);
          \draw[color=darkgrey,line width=1pt,->] (wt2) to (w3);
          \draw[color=darkgrey,line width=1pt,->] (wt2) to (w4);
          \draw[color=darkgrey,line width=1pt,->] (wt2) to (w5);
          \draw[color=darkgrey,line width=1pt,->] (wt2) to (w6);

          \begin{pgfonlayer}{background}
            \foreach \nodename in {w,w1,w2,w3,w4,w5,w6,wt2,a1,a2,a3,a4} {
              \coordinate (\nodename') at (\nodename);
            }
            \path[fill=grey1,draw=grey1,line width=1.1cm, line cap=round, line join=round] 
            (w') to (w1') to (w2') to (w3') to (w4')  to (w') to (a1') to (a2') to (a3') to (w') -- cycle;
            \path[fill=grey3,draw=grey3,line width=0.6cm, line cap=round, line join=round] 
            (wt2') to (w2') to (w3') to (w4') to (w5') to (w6')  to (wt2') to (a4') to (a3') to (a2') to (wt2') -- cycle;
          \end{pgfonlayer}
      \end{tikzpicture}};
      \node (table) at (1.0,0) {%
        \begin{minipage}{0.4\textwidth}
          \begin{eqnarray*}
            cite^{(ap)}(\{i_1,i_2\},\{j_2\}) & = & 1\\
            cite^{(ap)}(\{i_2,i_3\},\{j_2\}) & = & 2\\
            cite^{(aa)}(i_3,i_5) & = & 2 \\ 
            cite^{(aa)}(i_3,i_6) & = & 1  
          \end{eqnarray*}
        \end{minipage}
      };
    \end{tikzpicture}
  \end{center}
  \caption{\label{fig:attributes} Publication events of papers $j$ and $j'$ with their authors $i_1,\dots,i_4$ and references $j_1,\dots,j_6$. Additionally, authors ($i_5,i_6$) of some of the cited papers are given. Values of network attributes in this example are computed without any decay over time. Values for $cite^{(pp)}$ and $author$ are binary and are given by the lines (``edges'') connecting papers to papers, or authors to papers, respectively. Values of other network attributes on selected nodes are given on the righthand side.}
\end{figure}
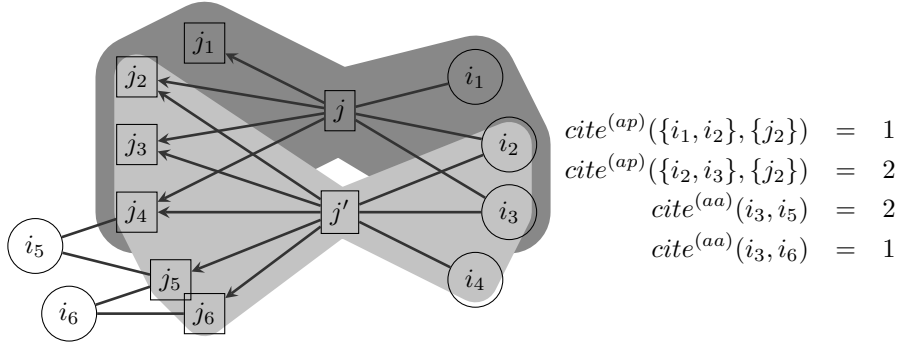

We illustrate the network attributes in Fig.~\ref{fig:attributes} in an example extending Fig.~\ref{fig:event}. The definition of model covariates is further illustrated with small stylized examples and graphics in Appendix~\ref{app:illustration}. 

\subsection{Summarizing history through time-varying network attributes}
\label{sec:attributes}

Given a sequence of publication events $E=\{(t_1,j_1,I_1,J_1),\dots,(t_n,j_n,I_n,J_n)\}$, the value of network attributes at time $t$ is a function of earlier events
\[
E_{<t}=\{(t_m,j_m,I_m,J_m)\in E\,;\;t_m<t\}\enspace,
\]
where we let the effect of past events decay over time with a given half-life period $T_{1/2}>0$ \citep{brandes2009networks,lerner2013modeling} using the following weight factor dependent on the elapsed time $t-t_m$
\[
w(t-t_m)=\exp\left(-[t-t_m]\frac{\log 2}{T_{1/2}}\right)\enspace.
\]
While using a decay is not theoretically necessary, it is empirically plausible that the effects of past collaborations or citations will fade over time.  The inclusion of a decay parameter typically leads to a better model fit \citep{lerner2023micro}.

The most important network attribute, which will be used as a basis for defining several covariates, encodes the extent to which a set of authors $I'$ has cited a set of papers $J'$ in joint publications. This attribute is a generalization of the ``sender-specific hyperedge degree'', defined in \citet{lerner2023relational}, to the case of two-mode hyperevent networks where the number of ``senders'' (authors in our case) is unconstrained. For a set of authors $I'\subseteq\mathcal{I}$ and a set of papers $J'\subseteq\mathcal{J}$, past author-to-paper citations are encoded in the attribute
\[
cite^{(ap)}_t(I',J')=\sum_{t_m<t}w(t-t_m)\cdot{\bf 1}(I'\subseteq I_m\wedge J'\subseteq J_m)\enspace.
\]
As special cases, this attribute already encodes to what extent a set of authors has coauthored papers, irrespective of the citations of their papers, by using $J'=\emptyset$ and it encodes the extent to which a set of papers have been cocited by other papers, regardless of the authors of these citing papers, by using $I'=\emptyset$.

For two papers $j,j'\in\mathcal{J}$, paper-to-paper citations are encoded in the attribute
\[
cite^{(pp)}_t(j,j')=\sum_{t_m<t}w(t-t_m)\cdot{\bf 1}(j=j_m\wedge j'\in J_m)\enspace.
\]
Note that at most one term of this sum is different from zero, since every paper is published only once, that is, appears only once as $j_m$ in a publication event $(t_m,j_m,I_m,J_m)$. 

For two authors $i,i'\in\mathcal{I}$, author-to-author citations are encoded in the attribute
\[
cite^{(aa)}_t(i,i')=\sum_{t_m<t}w(t-t_m)\cdot{\bf 1}(i\in I_m\wedge \exists t_{m'}\colon j_{m'}\in J_m\wedge i'\in I_{m'})\enspace.
\]
In words, the publication event of paper $j_m$ is counted in the sum given above if and only if $i$ is among the authors of $j_m$ and $j_m$ cites at least one paper $j_{m'}$ that has $i'$ among its authors.

For an author $i\in\mathcal{I}$ and a paper $j\in\mathcal{J}$, past author relations are encoded in the attribute
\[
author_t(i,j)=\sum_{t_m<t}w(t-t_m)\cdot{\bf 1}(j=j_m\wedge i\in I_m)\enspace.
\]
Again, at most one term of this sum is different from zero, since every paper is published only once.

For an author $i\in\mathcal{I}$ the \emph{author citation popularity} is the number of papers citing $i$ (downweighted by the elapsed time since the publication of the citing paper), defined in the node-level attribute 
\[
citepop^{(a)}_t(i)=\sum_{t_m<t}w(t-t_m)\cdot{\bf 1}(\exists t_{m'}\colon j_{m'}\in J_m\wedge i\in I_{m'})\enspace.
\]

The length of the reference list of a paper $j\in\mathcal{J}$ is captured by the following node-level attribute
\[
outdeg_t(j)=\sum_{t_m<t}w(t-t_m)\cdot{\bf 1}(j=j_m)\cdot|J_m|\enspace.
\]

We emphasize that many different ways to scale or normalize these attributes are possible. It is not in the scope of this paper to explore different variants -- or empirically to find the ``optimal'' normalization.

\subsection{Hyperedge covariates for network effects}
\label{sec:hyed_covs}

A generic covariate that can be specialized to several different types of covariates is called \emph{subset repetition of order $(k,\ell)$}. This covariate has already been defined in the preprint \citet{lerner2019rem} but has not been used in their empirical analysis. In our case, subset repetition formalizes three variants of effects: (1) if a set of authors $I'$ has already coauthored one or several papers (possibly together with others), they are likely to coauthor future papers (possibly together with other, potentially different authors); (2) if a set of papers $J'$ has already been co-cited (possibly within a larger list of references), they are likely to be co-cited again (possibly within a different larger list of references); (3) if a set of authors $I'$ have already coauthored a paper citing a set of references $J'$, they are likely to coauthor another paper, also citing the set of references $J'$. The crucial pattern in all of these variants is that some -- but not necessarily all -- of the nodes (authors and/or references) participating in a past event, participate again in a future event -- possibly within a varying set of further co-participating nodes. Subset repetition is parameterized by the sizes of the subsets that are to be repeated. For two non-negative integers $k$ and $\ell$ that are not both equal to zero, a set of authors $I\in\mathcal{I}$, and a set of papers $J\in\mathcal{J}$, subset repetition of order $(k,\ell)$ is defined as
\[
subrep_t^{(k,\ell)}(I,J)=\sum_{(I',J')\in{I\choose k}\times{J\choose \ell}}
\frac{cite^{(ap)}_t(I',J')}{{|I|\choose k}\cdot{|J|\choose \ell}}\enspace.
\]
The formula above iterates over all subsets of authors $I'\subseteq I$ and references $J'\subseteq J$ of the prescribed sizes and averages the past author-to-paper citation weight $cite^{(ap)}_t(I',J')$, defined above, over these combinations. Subset repetition can be specialized to the following three families of covariates. Recall that these and other covariates are further illustrated in Appendix~\ref{app:illustration}. 

\paragraph{Prior papers and prior joint papers}
For a set of authors $I\subseteq\mathcal{I}$ the average number of prior papers (downweighted by the elapsed time) is captured by the covariate
\[
\textup{prior papers}_t(I)=subrep_t^{(1,0)}(I,\emptyset)
\]
and is a measure of the past publication activity of authors in $I$. Previous collaboration among pairs of authors is captured by 
\[
\textup{prior joint papers}_t(I)=subrep_t^{(2,0)}(I,\emptyset)\enspace,
\]
which averages the number of coauthored papers (downweighted by the elapsed time) over all unordered pairs of authors in $I$. It would be possible to measure past collaboration among larger sets of authors via $subrep^{(k,0)}(I,\emptyset)$ for $k>2$. However, in our empirical analysis we found that the analyzed scientific network is too sparse to allow estimation of these effects, so that we dropped them from our models.\footnote{However, see \citet{lerner2023micro} who estimated subset repetition among coauthors up to order ten in a smaller and denser coauthoring network.}

A related covariate captures the heterogeneity of past publication activity via
\[
\textup{diff prior papers}_t(I)=\sum_{\{i,i'\}\in{I\choose 2}}
\frac{|cite^{(ap)}_t(\{i\},\emptyset)-cite^{(ap)}_t(\{i'\},\emptyset)|}{{|I|\choose 2}}\enspace.
\]
A positive parameter associated with this covariate would suggest that teams of coauthors are typically composed of authors having a larger dispersion in the number of prior publications than expected by chance alone. A negative parameter would suggest homogeneous teams, whose prior publication activity tends to be more similar than expected by chance alone.

\paragraph{Paper citation popularity and co-citation}
For a set of papers $J\subseteq\mathcal{J}$ the average number of past citations (downweighted by the elapsed time) is captured by the covariate
\[
\textup{paper citation popularity}_t(J)=subrep_t^{(0,1)}(\emptyset,J)\enspace.
\]
A positive parameter associated with this covariate would indicate a ``rich get richer'' effect in the sense that papers with more past citations are also cited at a higher rate in the future. Previous co-citations of pairs and triples of papers are captured by
\begin{eqnarray*}
\textup{paper-pair cocitation}_t(J)&=&subrep_t^{(0,2)}(\emptyset,J)\\
\textup{paper-triple cocitation}_t(J)&=&subrep_t^{(0,3)}(\emptyset,J)\enspace.
\end{eqnarray*}
Positive parameters for these covariates would suggest that pairs (or triples) of papers that jointly appeared in reference lists are more likely to be cocited again in future publications.

\paragraph{Author citation repetition}
The hypothetical tendency of (groups of) authors to repeatedly cite the same papers (compare Fig.~\ref{fig:event}) is captured by the following covariates combining the history of a set of authors $I\subseteq\mathcal{I}$ and a set of papers (potential references) $J\subseteq\mathcal{J}$
\begin{eqnarray*}
\textup{author citation repetition}_t(I,J)&=&subrep_t^{(1,1)}(I,J)\\
\textup{author-pair citation repetition}_t(I,J)&=&subrep_t^{(2,1)}(I,J)\\
\textup{author-triple citation repetition}_t(I,J)&=&subrep_t^{(3,1)}(I,J)\enspace.
\end{eqnarray*}

\paragraph{Citing a paper and its references}
The tendency to adopt (some of) the references of a cited paper -- or, conversely, to cite a paper and some of its citing papers -- is captured by the covariate measuring the past citation density within a set of papers $J\subseteq\mathcal{J}$ (representing a possible list of references):
\[
\textup{cite paper and its refs}_t(J)=\sum_{\{j,j'\}\in {J \choose 2}}
\frac{cite^{(pp)}_t(j,j')+cite^{(pp)}_t(j',j)}{\binom{|J|}{2}}\enspace.
\]
We note that this covariate is likely to have an influence on the citation disruption index \citep{park2023papers} since a positive parameter would suggest a lower expected CDI.

\paragraph{Paper outdegree popularity}
We define the average length of reference lists of a set of papers $J\subseteq\mathcal{J}$ via the covariate
\[
\textup{paper outdegree popularity}_t(J)=\sum_{j\in J}
\frac{outdeg_t(j)}{|J|}\enspace.
\]
A positive parameter associated with this covariate would suggest a tendency to cite papers with long reference lists. This covariate can be seen as a precondition of the above covariate ``citing a paper and its references'' which should be included to control for varying length of reference lists.

\paragraph{Collaborate with citing author}
The tendency of scientists to coauthor papers with those who cited their previous work is captured by the covariate measuring the past citation density within a set of authors $I\subseteq\mathcal{I}$:
\[
\textup{collab w citing author}_t(I)=\sum_{\{i,i'\}\in {I \choose 2}}
\frac{cite^{(aa)}_t(i,i')+cite^{(aa)}_t(i',i)}{\binom{|I|}{2}}\enspace.
\]

\paragraph{Author citation popularity}
We define the average citation popularity of a set of authors $I\subseteq\mathcal{I}$ via the covariate
\[
\textup{author citation popularity}_t(I)=\sum_{i\in I}
\frac{citepop^{(a)}_t(i)}{|I|}\enspace,
\]
and the heterogeneity in author citation popularity (similar to the heterogeneity in the past publication activity) via the covariate
\[
\textup{diff auth cite pop}_t(I)=\sum_{\{i,i'\}\in{I\choose 2}}
\frac{|citepop^{(a)}_t(i)-citepop^{(a)}_t(i')|}{{|I|\choose 2}}\enspace.
\]
A positive parameter associated with the author citation popularity would suggest that authors whose papers have been cited more often in the past are more likely to publish future papers. A positive parameter associated with the difference in authors' citation popularity would suggest that teams of coauthors are typically composed of authors with varying citation popularity, e.\,g., senior scientists coauthoring together with junior scientists.

\paragraph{Author self-citation}
Capturing the tendency of authors to cite their own past work, we define the covariate \emph{self-citation} measuring the density of the two-mode subgraph connecting a set of authors $I\subseteq\mathcal{I}$ and a set of papers $J\subseteq\mathcal{J}$ with respect to the $author$ relation
\[
\textup{self-citation}_t(I,J)=\sum_{i\in I,\;j\in J}
\frac{author_t(i,j)}{|I|\cdot |J|}\enspace.
\]

\paragraph{Triadic closure (author pairs)} The remaining list of covariates are variants of triadic closure in the sense that pairs of nodes (which can be authors or papers) are characterized by their past relations to ``third'' nodes (which, again, can be authors or papers) in various types of relations, defined as network attributes above. The first four covariates in this family establish indirect relations among pairs of authors, that may have an influence on the probability that these authors collaborate in the future.

\paragraph{Coauthor closure} The covariate \emph{common coauthor} measures to what extent authors in $I\subseteq\mathcal{I}$ have coauthored papers with the same ``third'' authors. To abbreviate notation, we write $coauth_t(i,i')=cite^{(ap)}_t(\{i,i'\},\emptyset)$.
\[
\textup{common coauthor}_t(I)=\sum_{\{i,i'\}\in{I\choose 2}\wedge i''\neq i,i'}
\frac{\min[coauth_t(i,i''),coauth_t(i',i'')]}{{|I|\choose 2}}\enspace.
\]
In the summation above, the ``third'' author index $i''$ runs over all authors $i''\in\mathcal{I}$, different from the two focal authors $i$ and $i'$.

\paragraph{Citing common paper}
A related covariate \emph{citing common paper} measures to what extent authors in $I\subseteq\mathcal{I}$ have cited the same papers in the past.
\[
\textup{cite common paper}_t(I)=\sum_{\{i,i'\}\in{I\choose 2}\wedge j\in\mathcal{J}}
\frac{\min[cite^{(ap)}_t(\{i\},\{j\}),cite^{(ap)}_t(\{i'\},\{j\})]}{{|I|\choose 2}}\enspace.
\]

\paragraph{Citing common author}
The covariate \emph{citing common author} measures to what extent authors in $I\subseteq\mathcal{I}$ have cited (papers published by) the same authors in the past.
\[
\textup{cite common auth}_t(I)=\sum_{\{i,i'\}\in{I\choose 2}\wedge i''\neq i,i'}
\frac{\min[cite^{(aa)}_t(i,i''),cite^{(aa)}_t(i',i'')]}{{|I|\choose 2}}\enspace.
\]

\paragraph{Cited by common author}
Finally, the covariate \emph{cited by common author} reverses these relations and measures to what extent (papers published by) authors in $I\subseteq\mathcal{I}$ have been cited by the same authors in the past.
\[
\textup{cited by common auth}_t(I)=\sum_{\{i,i'\}\in{I\choose 2}\wedge i''\neq i,i'}
\frac{\min[cite^{(aa)}_t(i'',i),cite^{(aa)}_t(i'',i')]}{{|I|\choose 2}}\enspace.
\]

\paragraph{Triadic closure (author-paper pairs)} The remaining four ``triadic closure'' covariates establish indirect relations between authors and papers via various types of relations.

\paragraph{Adopt citations of coauthors}
A tendency of authors to cite the papers that have been cited in the past by their past coauthors can be assessed by the following covariate
\[
\textup{adopt cite coauth}_t(I,J)=\sum_{i\in I\wedge j\in J \wedge i'\neq i}
\frac{\min[coauth_t(i,i'),cite^{(ap)}_t(\{i'\},\{j\})]}{|I| \cdot |J|}\enspace.
\]

\paragraph{Citing papers of coauthors}
Similarly, a tendency of authors to cite the papers that have been published by their past coauthors can be assessed by the following covariate
\[
\textup{cite coauthor's paper}_t(I,J)=\sum_{i\in I\wedge j\in J \wedge i'\neq i}
\frac{\min[coauth_t(i,i'),author_t(i',j)]}{|I| \cdot |J|}\enspace.
\]

\paragraph{Citation-repetition on the author level}
We define a covariate \emph{author-author citation repetition} that can test the tendency of authors to repeatedly cite the papers of the same authors
\[
\textup{auth-auth cite repet}_t(I,J)=\sum_{i\in I\wedge j\in J \wedge i'\neq i}
\frac{\min[cite^{(aa)}_t(i,i'),author_t(i',j)]}{|I| \cdot |J|}\enspace.
\]
Note that this covariate is structurally different from ``author citation repetition'' since the latter tests whether the same author tends to repeatedly cite the same paper, while the new ``author-author citation repetition'' tests whether the same author $i$ tends to repeatedly cite potentially different papers written by the same other author $i'$.

\paragraph{Citation-reciprocation on the author level}
Reversing a relation in the previously defined author-author citation repetition, we define a covariate \emph{author-author citation reciprocation} testing the tendency of authors to cite the papers of those other authors who have previously cited their papers
\[
\textup{auth-auth cite recipr}_t(I,J)=\sum_{i\in I\wedge j\in J \wedge i'\neq i}
\frac{\min[cite^{(aa)}_t(i',i),author_t(i',j)]}{|I| \cdot |J|}\enspace.
\]

We emphasize that many different ways to scale or normalize these covariates are possible and there are also many possibilities to define yet other structurally different covariates. The objective of this paper is to define a representative set of covariates, illustrating structurally different ways how past publication events may shape the distribution of future publications. It is impossible -- and not in the scope of this paper -- to provide a ``complete'' list of covariates.

\subsection{Hyperedge covariates dependent on exogenous factors}

Another family of hyperedge covariates may be expressed in terms of numeric or categorical exogenous attributes of authors or papers. Examples of such exogenous attributes for papers include subject classifiers, keywords, and numeric summaries of the text of abstracts or whole papers. For the case of authors, exogenous attributes include gender, seniority, or affiliation. Types of hyperedge covariates based on such exogenous attributes include summary statistics of attribute values over authors, papers, or of the interaction between author attributes and paper attributes (e.\,g., interaction between research interest of authors and subject classifiers of papers cited by them). Simple examples for the functional form of such covariates are provided by the two effects ``author citation popularity'' and ``diff author citation popularity'', defined above. Although these covariates are defined as functions of the history-dependent node-level attribute $citepop^{(a)}_t(i')$, their functional form could be re-used to define similar covariates dependent on exogenous attributes (e.\,g., seniority or gender).

Since we do not use any exogenous attributes in our empirical example below -- and since we believe that their definition is straightforward and simple, compared to some of the more complex history-dependent attributes -- we do not explicitly define covariates depending on exogenous attributes. We further refer the reader to \citet{lerner2023relational} for explicit examples of such covariates.

\subsection{Hyperedge covariates dependent on the published paper}

We note that none of the covariates defined above takes the focal paper $j_m$ of a publication event $(t_m,j_m,I_m,J_m)$ into account. Indeed, since we defined only covariates dependent on the history of the process, considering the paper $j_m$ would not make any difference, since a paper has an empty history at its publication time. If, however, we considered covariates dependent on exogenous attributes, it would be possible in a straightforward manner to also take attributes of $j_m$ into account. For instance, overlap in keywords of $j_m$ with the papers in $J_m$, related subject classifiers, or some measure of text similarity, could make it more likely that $j_m$ has $J_m$ as its reference list.

\section{Empirical example}
\label{sec:example}

We establish the empirical value of the new RHEM we have described in an analysis of a large scientific networks of bibliographic citations. The analysis focuses on the estimation and interpretation of the mixed-mode effects, and on the assessment of their relative importance vis-\`a-vis the set of more common ``network effects'' routinely incorporated in empirical specification of statistical models for citation networks \citep{newman2004coauthorship} to capture  indegrees and outdegrees and clustering of authors and papers.    

\subsection{Data}
\label{sec:data}

\begin{figure}
  \centering
  \includegraphics[width=0.8\textwidth]{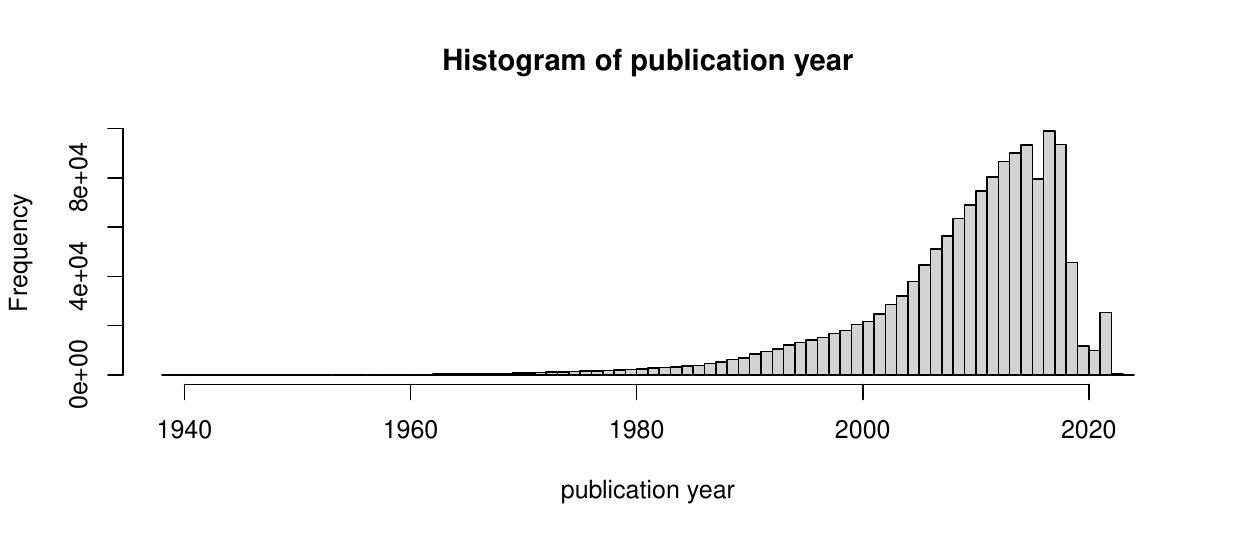}
  \caption{\label{fig:hist_year} Histogram of publication year of papers used in the empirical analysis.}
\end{figure}

To construct our data, we take the Aminer Citation Network Dataset (\url{https://www.aminer.org/citation}) ``DBLP-Citation-network V14'' \citep{tang2008arnetminer}, which we restrict to journal publications (indicated by the field value \texttt{doc\_type="Journal"}) and we also retain only those references that appear in this restricted data (i.\,e., only citations from and to journal articles). If by this process, the list of references of some paper $j$ becomes empty, we drop the respective publication event from our analysis (however, $j$ could still be cited by another paper in our data). The resulting 1,416,353 papers (i.\,e., publication events) are written by 1,286,941 unique authors. The number of authors per paper ranges from one to 134 with mean equal to 2.8 and the number of references (included in the analysis) ranges from 1 to 2,268 with the mean number of references equal to 8.6. The year of publication ranges from 1939 to 2023, where the bulk of papers has been published after the year 2000; see the distribution of publication years in Fig.~\ref{fig:hist_year}.

\subsection{Preprocessing, estimation, and reproducibility}
\label{sec:implementation}

Explanatory variables (i.\,e., hyperedge covariates) are computed with the open-source software \texttt{eventnet} \citep{lerner2020reliability,lerner2021dynamic}, available at \url{https://github.com/juergenlerner/eventnet}. The \texttt{eventnet} software samples $q$ non-event hyperedges associated with each observed hyperevent (i.\,e., publication event) and computes all covariates of observed events and sampled non-events. For defining the set of authors $\mathcal{I}_t$ that could publish a paper in year $t$ and the set of papers $\mathcal{J}_t$ that could be cited in year $t$ we use all authors, respectively papers, that appear in any publication event at $t$ or earlier. In particular, we do never remove authors from $\mathcal{I}_t$ since we do not have any information about when an author cannot (or decides not to) publish anymore. (Given such information, \texttt{eventnet} would allow to remove nodes from the risk set.) Note however that covariates capturing author activity (e.\,g., ``prior papers'') have a decay and approach zero after longer periods of inactivity. Thereby the models could learn implicitly that authors are unlikely to publish again after longer periods of inactivity.

We set the number of non-events per event to $q=5$. All network attributes are defined with the half life period $T_{1/2}$ set to three years. We apply to all covariates the square-root transformation $x\mapsto\sqrt{x}$, which consistently results in a better fit for all models. We note that a general search for optimal $q$, half life periods, or variable transformation is out of scope of this paper. However, we comment more on the sample size and choice of the half-life parameter after having sketched the estimation method.

For each model we estimate parameters with the \texttt{coxph} function in the \texttt{R} package \texttt{survival} \citep{therneau2013modeling,therneau2015survival}, using robust estimation. As noted when defining the intensity (\ref{eq:lambda}) and the (sampled) likelihood function, (\ref{eq:likelihood}) and (\ref{eq:likelihood_sampled}), our models are stratified CoxPH models where the strata are defined by the combination of the number of authors and the number of references. However, for reasons lying in the implementation of the \texttt{coxph} function, we must explicitly provide the information that only controls and events having the same event time (i.\,e., publication time in our case) are put in the same fraction of the sampled likelihood (\ref{eq:likelihood_sampled}). This has for instance been noted by \citet{aalen2008survival}, who write when discussing the Cox model with nested case-control sampling: ``\emph{For computing, one may use standard software for Cox regression and other relative risk regression models, formally treating the label of the sampled risk sets as a stratification variable in the model [\dots]} \citep[][p.~197]{aalen2008survival}''. Thus, in our case, the \texttt{strata} argument of the \texttt{coxph} function is a variable defined by the concatenation of publication time (year), number of authors, and number of references.

We observe that robust estimation via \texttt{coxph} leads for some of the effects to considerably different estimates for standard errors than default estimation. However, this rarely affects significance at conventional levels ($p<0.01$ or $p<0.001$), most likely due to the size of our sample. We nevertheless recommend robust estimation in studies where it is reasonable to expect the presence of residual dependencies among the observations.

Since, in general, many papers are published in a given year, we have a data set with ``tied event times'' \citep{kalbfleisch1973marginal,breslow1974covariance,efron1977efficiency,hertz1997validity}. In the \texttt{coxph} function that we adopt tied even times are handled by default through the so-called ``Efron approximation'' \citep{efron1977efficiency, hertz1997validity}. We further note that, even in the presence of tied event times, the assumption that the values of hyperedge covariates at time $t$ are only functions of prior publication events at time $t'<t$ still holds. Our models assume that simultaneous events are conditionally independent of each other, given the set of previous events \citep{lerner2013conditional}. In practice, this implies that, for instance, authors do not base their decision about the selection of coauthors or references on other publications appearing in the same year. While testing the implications of this assumption is outside of scope of this paper, we do not think that this assumption could be considered implausible or unnecessarily restrictive. We note that, for instance, patent collaboration and citation networks typically come with a much finer time granularity, often given by the day. Scientific papers published in the last few years often provide more fine-grained information on (online) publication dates given by the day. When available, our model is able to make efficient use of these pieces of finer-grained information about the timing of events.

To assess the reliability of parameter estimates with the given chosen value of $q$, we generate 100 bootstrap samples \citep{efron1992bootstrap} -- a possibility mentioned by \citet{lerner2020reliability} -- and re-estimate parameters of the ``joint model'' from Table~\ref{table:coefficients} on each of these samples. We find that for each of the 20 significant effects, all 100 parameters estimated by bootstrap sampling have the same sign. For the single non-significant effect (``citing common author''), the parameters estimated by bootstrap sampling are rather centered around zero -- coinciding with the finding that this parameter is not significantly different from zero. Thus, judging by bootstrap sampling, our sample size seems to be large enough to get reliable estimates.

We argue -- already for theoretical reasons -- that the choice of the half-life period is likely to impact findings. Changing the half-life period implies that we test different effects: long-term effects of, e.\,g., past collaboration or citation versus their short-term effects. It could be expected that long-term effects are different from short-term effects, at least for some of the structural patterns. For instance, there might be an effect if two scientists have ever co-authored a paper, even if this happened decades ago; but this long-term effect might be different from a short-term effect of having co-authored a paper few years ago. To probe this issue further, we also fitted the three models from Table~\ref{table:coefficients} where all covariates were computed without any decay (``infinite'' half-life). Indeed findings are qualitatively different. One notable difference is that in the no-decay model, the parameter of ``prior papers'' (average number of prior papers of the authors included in a hyperedge) has a significantly negative sign, whereas it is positive in the models reported in Table~\ref{table:coefficients} (with half-life set to 3 years). This could be explained by the fact that we are analyzing a period of more than 60 years. Authors who have been very active in the (distant) past may simply be no longer active scientists and therefore do no longer publish at all. The model without decay might seek to explain this by a negative effect of the number of prior papers (e.\,g., once scientists have accumulated many citations, they are closer to retirement age and stop publishing). More generally, models without decay are not able to express that scientists may be active in some periods but later become less active, that collaborations among scientists might dissolve, that once popular papers might lose their popularity (and hence get cited less), and so on. While we discard the model without decay out of theoretical reasoning, we argue that it is hard to find optimal half-life parameters -- at least for a network of this size. The optimal value might vary by effect and models might even be specified by adding the same structural effect several times with varying half-life parameters.

The analysis we present next is fully reproducible. We provide all software and scripts (preprocessing, eventnet configuration files, and R scripts), as well as explanations on how to use them, on a dedicated page in the \texttt{eventnet} wiki, available at: \url{https://github.com/juergenlerner/eventnet/wiki/Coevolution-of-collaboration-and-references-to-prior-work-(tutorial)}. We further note that \texttt{eventnet} is open source published under the GPL 3.0 license, allowing inspection, modification, and re-publishing of code and providing a reference implementation for computing hyperedge covariates.

\subsection{Results}
\label{sec:results}

\paragraph{Model parameters}
Table~\ref{table:coefficients} provides estimated parameters of the coauthor model, citation model, and joint model. Note that, while the observed events are identical for the three models, the sampled non-events are different and are indeed sampled from different spaces: sets of authors, sets of papers, and pairs of author sets and paper sets, respectively.

\begin{table}
\caption{Estimated parameters and standard errors (in brackets) obtained by robust estimation.}
\label{table:coefficients}
\centering{\footnotesize
\begin{tabular}{l r r r}
\hline
 & coauthors & citations & joint \\
\hline
prior papers     & $0.215 \; (0.008)^{***}$  &                           & $0.129 \; (0.004)^{***}$  \\
diff prior papers    & $0.181 \; (0.004)^{***}$  &                           & $0.185 \; (0.003)^{***}$  \\
prior joint papers                    & $1.729 \; (0.018)^{***}$  &                           & $0.823 \; (0.009)^{***}$  \\
author citation popularity      & $-0.032 \; (0.002)^{***}$ &                           & $-0.017 \; (0.001)^{***}$ \\
diff auth cite pop     & $0.027 \; (0.002)^{***}$  &                           & $-0.007 \; (0.001)^{***}$ \\
collab w citing author      & $0.503 \; (0.013)^{***}$  &                           & $0.318 \; (0.006)^{***}$  \\
common coauthor          & $-0.080 \; (0.004)^{***}$ &                           & $-0.014 \; (0.003)^{***}$ \\
citing common paper    & $-0.008 \; (0.006)\phantom{^{***}}$       &                           & $-0.112 \; (0.004)^{***}$ \\
citing common author   & $-0.056 \; (0.003)^{***}$ &                           & $0.001 \; (0.002)\phantom{^{***}}$        \\
cited by common author & $-0.059 \; (0.002)^{***}$ &                           & $0.007 \; (0.002)^{***}$  \\
paper citation popularity                 &                           & $0.188 \; (0.001)^{***}$  & $0.170 \; (0.001)^{***}$  \\
paper-pair cocitation                 &                           & $0.020 \; (0.011)\phantom{^{***}}$        & $0.032 \; (0.012)^{**\phantom{*}}$   \\
paper-triple cocitation                 &                           & $-0.208 \; (0.021)^{***}$ & $-0.170 \; (0.021)^{***}$ \\
author citation repetition          &                           & $0.829 \; (0.018)^{***}$  & $0.643 \; (0.017)^{***}$  \\
author-pair citation repet           &                           & $-0.524 \; (0.016)^{***}$ &                           \\
author-triple citation repet          &                           & $-0.195 \; (0.020)^{***}$ &                           \\
paper outdegree popularity          &                           & $0.192 \; (0.005)^{***}$  & $0.195 \; (0.006)^{***}$  \\
cite paper and its refs     &                           & $7.746 \; (0.023)^{***}$  & $7.360 \; (0.024)^{***}$  \\
adopt citation of coauthor         &                           & $-0.038 \; (0.006)^{***}$ & $-0.264 \; (0.007)^{***}$ \\
self citation                &                           & $1.745 \; (0.028)^{***}$  & $1.799 \; (0.025)^{***}$  \\
cite coauthor's paper              &                           & $0.497 \; (0.017)^{***}$  & $0.291 \; (0.016)^{***}$  \\
author-author citation repet          &                           & $-0.042 \; (0.017)^{*\phantom{**}}$   & $-0.381 \; (0.015)^{***}$ \\
author-author citation recip       &                           & $-0.258 \; (0.014)^{***}$ & $-0.343 \; (0.014)^{***}$ \\
\hline
AIC                                 & $20602304.582$            & $19168404.629$            & $18927741.318$            \\
Num. events                         & $1,416,353$                 & $1,416,353$                 & $1,416,353$                 \\
Num. obs.                           & $8,493,351$                 & $8,493,344$                 & $8,493,353$                 \\
\hline
\multicolumn{4}{l}{\scriptsize{$^{***}p<0.001$; $^{**}p<0.01$; $^{*}p<0.05$}}
\end{tabular}}
\end{table}

Parameters typically -- but not always -- have the same sign and significance levels in the joint model and in one of the two submodels. An exception is the difference in the author citation popularity (``diff auth cite pop'') and the effect assessing whether two scientists are likely to become coauthors if they have been ``cited by [a] common author''. Below we discuss findings related to some of the strongest, or most interesting and interpretable effects.

Findings explaining the formation of coauthor teams include a positive effect of the number of prior papers (i.\,e., scientists  publishing more in the recent past, tend to publish at a higher rate in the future) and a positive effect of the number of prior joint papers (suggesting a tendency to repeat scientific collaboration with the same coauthors). These self-reinforcing process are consistent with established hypotheses on the cumulative advantage in science \citep{price1976general}, and social inertia in scientific collaboration \citep{ramasco2006social}, respectively, 

The difference in the number of prior papers has a positive effect, suggesting that coauthor teams are typically composed of scientists with varying publication history, for instance, senior scientists who publish papers together with their PhD students. We find that scientists have a tendency to collaborate with those who cited their papers. The four triadic effects explaining coauthoring, from ``common coauthor'' to ``cited by common author'', are mostly negative in the coauthoring model, with the effect of having cited common papers not being significant. This result is in line with previous findings on negative closure in coauthor networks \citep{lerner2023micro} and hyperevent networks of meeting events \citep{lerner2021dynamic,lerner2022dynamic}. Following the interpretation given in these papers, negative closure suggests the existence of various dense groups of actors that do not merge over time -- or, alternatively, the existence of actors occupying stable broker positions, surrounded by ``structural holes,'' or open triads \citep{burt1992structural}.

Strong and positive effects explaining the selection of references include the number of times that these references have been cited before (paper citation popularity), the tendency of authors to repeatedly cite the same papers (author citation repetition), to cite their own papers (self-citation), and to cite papers published by their past coauthors (cite coauthor's paper). Another strong positive effect is the tendency to cite papers together with some of their references. This is consistent with the finding by \citet{park2023papers} that the citation disruption index is lower than expected in a randomized citation network (although it does not say anything about their main claim stating that the CDI is declining over time). 

\paragraph{Contributions of covariates in the joint model}

Table~\ref{table:contributions_joint} presents contributions of the individual covariates to the log likelihood of the joint model. The first numeric column gives the differences in log likelihoods of the respective single-parameter model (specified with only one covariate at a time) over the null model (no parameters). The second numeric column gives the differences between the ``full model'' (specified with all 21 covariates) and the models with 20 parameters obtained by dropping the respective effect from the full model. The last column gives the type of effect: ``A'' for pure authorship effects, that is, covariates that could still be computed if we discarded the set of cited papers $J_m$ from all publication events $(t_m,j_m,I_m,J_m)$, ``C'' for pure citation network effects, that is, covariates that could still be computed if we discarded the set of authors $I_m$ from all publication events, and ``M'' for mixed effects that depend on authors and citations. The log likelihood of the null model is -11,063,051 and the difference in log likelihood between the full model and the null model is 1,599,202.

\begin{table}
\caption{Contributions to the log likelihood of individual covariates in the joint model. First numeric column: improvements over the null model; second numeric column: contributions in the full model. Last column: type of effect (``A'' for pure coauthorship effects, ``C'' for pure citation effects, ``M'' for mixed effects).}
\label{table:contributions_joint}
\centering
\begin{tabular}{rrrc}
  \hline
 & to null & in full & type\\ 
  \hline
cite paper and its refs & 1,150,680 & 237,136 & C \\ 
  paper citation popularity & 629,911 & 115,535 & C \\ 
  author citation repetition & 614,921 & 3,495 & M \\ 
  self citation & 537,228 & 9,967 & M \\ 
  num prior joint papers & 508,556 & 21,586 & A \\ 
  num prior papers & 508,359 & 1,230 & A \\ 
  author-author citation repetition & 441,856 & 1,545 & M \\ 
  collab w citing author & 415,377 & 4,324 & M \\ 
  diff num prior papers & 402,533 & 5,537 & A \\ 
  paper-pair cocitation & 400,152 & 278 & C \\ 
  author-author citation reciprocation & 359,472 & 1,331 & M \\ 
  cite coauthor's paper & 355,529 & 849 & M \\ 
  adopt citation of coauthor & 329,170 & 4,649 & M \\ 
  author citation popularity & 251,639 & 198 & M \\ 
  diff auth cite pop & 217,229 & 55 & M \\ 
  paper-triple cocitation & 212,057 & 1,123 & C \\ 
  citing common paper & 202,071 & 1,682 & M \\ 
  common coauthor & 189,043 & 67 & A \\ 
  paper outdegree popularity & 168,659 & 12,428 & C \\ 
  citing common author & 160,407 & 0 & M \\ 
  cited by common author & 112,673 & 66 & M \\
   \hline
\end{tabular}
\end{table}

We find that the strongest effect, by far, is the tendency of authors to cite a paper together with some of the references of that cited paper. The implied pattern in citation networks emerges if, for instance, paper $j_1$ is cited by $j_2$ and later paper $j_3$ cites $j_1$ and $j_2$. Different explanations could lead to this pattern: authors may copy part of the references of a paper they cite -- or authors may search for citations to a paper they want to cite, and include also some of these citing papers in their reference list. Besides these purely mechanistic explanations, the papers' content (similarity) or relevance for a specific new paper could as well explain this pattern. If $j_2$ cites $j_1$, it is a signal that these two papers are likely to be related; if $j_3$ cites $j_2$ they are related as well; by transitivity $j_1$ and $j_3$ are related, which may increase the probability that $j_3$ cites $j_1$. With the given data we cannot distinguish between such different explanations for the same structural effect.

The second-strongest effect is the paper citation popularity -- or ``Matthew effect''  \citep{merton1968matthew} --  indicating that papers that have been cited more often in the past accumulate citations at a higher rate in the future. These two covariates are pure citation network effects and they have the strongest individual contributions over the null model and in the full model.

The next effects, ordered by contributions over the null model, are mixed effects: the tendency of authors to repeatedly cite the same papers and the tendency of authors to cite their own prior work. For these effects we observe that the contributions over the null model can lead to a different relative ordering than the contributions in the full model. A possible explanation is that for some effects we have included more related effects than for others -- which implies that the focal effect has a smaller additional contribution in the full model. For instance, part of the contributions of ``author citation repetition'' can presumably be explained by the papers' citation popularity and/or by the authors' publication activity (``prior papers''). Conversely, the ``paper outdegree popularity'' has a relatively weak effect over the null model but the fourth strongest effect in the full model. We note that the outdegree popularity is one out of few effects that reverse the roles of citing and cited papers: papers that cite many others (high outdegree) are more likely to be cited. The only other effect that also reverses the roles of ``citing'' and ``being cited'' is the ``author-author citation reciprocation'' -- although in this effect the nodes are authors and not papers.

The effects at Position 5 and 6 are the pure authorship effects ``prior joint papers'' (that is, repeatedly coauthoring papers with the same others) and ``prior papers'' (that is, past publication activity measured by the number of prior papers, downscaled by the elapsed time).
Some of the weakest effects are the four triadic closure effects that relate two authors through past relations to common cited papers, common coauthors, common cited authors, or by having been cited by common authors.
The difference in contributions from one effect to the next, ordered by contributions over the null model, are associated with relatively small jumps in the log-likelihood function -- with the notable exception of the unexpectedly strong contribution of the strongest effect (partial copying of the reference lists of cited papers).

\section{Conclusion}
\label{sec:conclusion}

We proposed, implemented, and tested new hyperedge covariates to extend RHEM to the analysis of scientific networks -- dynamic mixed two-mode networks whose nodes are authors and papers, where authors are connected to the papers they write, and papers are connected to other papers they cite. The models presented in this paper extend the analytical possibilities in the study of scientific citation networks by affording consideration of a broader set of dynamic network dependencies. Existing models considering only dependencies in either the coauthoring network, or in the citation network, respectively, can only take into account a subset of possible dependencies in complex systems of scientific citation networks. 

We argued and demonstrated that the network dynamics of the two components of this system (authors-by-papers, and papers-by-references) are heavily interdependent. On the one hand, the selection of reference lists depends on the authors of the current and past papers. On the other, the formation of teams of coauthors also depends on their and others past citations. We argued, further, that interaction in scientific networks is inherently polyadic, since a single publication event connects a varying and theoretically unbounded number of authors and cited papers. RHEM are a family of statistical models that can deal with the polyadic nature of scientific networks, and the main contribution of this paper is to adapt RHEM to time-stamped mixed two-mode networks \citep{snijders2013model}.

We adapted RHEM to networks of publication events, defining a list of representative covariates that implement structurally different effects, and conducted an empirical analysis of a large scientific network. We found that mixed-mode effects -- explaining scientific publications through a combination of authoring and citation relations -- are generally strong and can be used to test theoretically relevant and empirically meaningful effects explaining scientific collaboration and research orientation. Likewise, we found significant  polyadic effects -- acknowledging that co-citation probabilities of two or more papers are not determined by individual citation probabilities of single papers. Most notably, we found a significant tendency to cite a paper together with parts of its references. The corresponding effects are among the strongest in our models. We think that this result is particularly important because it suggests that the impact of an individual paper depends, at least in part, on the propensity of that paper to become part of a ``package'' of other papers that are frequently cited together. This seems to be a particularly important conjecture in the light of the standard practice to consider the number of citations received as a general measure of scientific impact \citep{radicchi2008universality}.

We emphasize that the contribution of our paper is neither in the general form of the likelihood function, nor the estimation procedure as both should be considered well established features of relational event and hyperevent models \citep{bianchi2023relational, butts2008relational, perry2013point,lerner2023relational}. Our emphasis on dependence among publications determined by intersecting references is also not new.  The germane concept of  bibliographic coupling as an approach to measure similarity among scientific papers is well-established within bibliometry at least since the work of \citet{kessler1963bibliographic}. Rather, the novelty of the present work should be evaluated in terms of the new effects (hyperedge covariates) for publication events that can be tested after inclusion in the partial likelihood (\ref{eq:likelihood}) and (\ref{eq:likelihood_sampled}). These effects allow us to go well beyond aggregate measures of similarity among publications that may be derived from bibliographic coupling \citep{liu2017new}. The new effects that we have presented in this paper allow to specify and test the possible network micro-mechanisms underlying observed macrolevel patterns of bibliographic coupling. The empirically validated conclusion that some of these new effects make significant and very strong contributions to our understanding of the structure of scientific networks and their constructive relational mechanisms may be interpreted as evidence that models not accounting for the polyadic nature of publication events, and/or not accounting for the interdependence between authoring and citation networks might miss important dependencies of current publications on prior publications. Thus, our paper makes an important step forward in the quest towards more empirically plausible and informed models for citation networks. 

Much work remains to do be done to alleviate some of the limitations of the study. For example, a comprehensive approach to evaluating the goodness of fit of the models we have presented like the one recently established by \citet{amati2024goodness} is still not available at a scale comparable to that of the sample we have analyzed. Also, model estimation procedures based on the sampled partial likelihood that we have implemented building on prior studies \citep{perry2013point} assumes that events are conditionally independent, given the earlier events, and that this dependence is fully reflected in the model specification.  We have not tested for potential violations of this assumption. We simply followed best recommended practice \citep{aalen2008survival}  and controlled for generic sources of violation by adopting robust estimation techniques. A final issue worth mentioning is more of an empirical issue than a limitation of the model. In studies based on case control sampling, like ours, it is common  for a particular covariate (or set of covariates) to be particularly sparse in the risk set but not infrequent in the subset of observed cases. \citet{lerner2020reliability} discuss this problem of sparsity and the convergence issues that it may generate in unfavorable empirical circumstances.   

Despite these actual and potential limitations, we think our study provides believable evidence that citations linked to scientific papers are not independent: citations reported in a paper are typically overlapping with the citations reported in cited papers. The fact that a paper citing another often also cites some of its references may not be surprising in light of what we now know about the practice of scientific production \citep{meng2023hidden, pandey2020analysis, simkin2002read}. In fact, we would expect that technological progress in information retrieval and the increased availability of large digital paper repositories will be likely to accentuate these trends. What is surprising is that, before the current paper, no model was available to adequately incorporate this predictable empirical feature in the analysis of scientific citation networks. 

We think that the results presented -- while only suggestive at the current stage -- have potentially far-reaching implications for the evaluation of scientific productivity, science policy and, ultimately, for the evaluation of scientists' academic career.  If citations are not independent, but -- as our study suggests -- ``come in packages,'' then it should be obvious that bibliometric indexes computed under the assumption of independence will be imperfect, and possibly inflated by the tendency of specific references to be included in subsets of references that are frequently cited together by the same authors, their coauthors or simply by other papers connected by direct citation links.  Understandably perhaps, individual success measured by number of citations remains the dominant metric for assessing scientific impact. Our work suggests that individual success may be affected by non-individual components generated by complex network dependencies.

Some of the effects that we have defined in this paper will help derive and test specific hypotheses about the mechanisms that regulate the size and composition of ``citation packages,''  the frequency with which they recur across papers, and the tendency of authors and coauthors to include citations in packages. In the light of the results presented in this paper, we believe that future research in the science of science should re-examine the implicit assumption found in many studies that the number of citations received represents a direct measure of scientific impact at the individual, or even team-level \citep{fortunato2018science}.

In closing, it is worth emphasizing that the dynamic structure of scientific citation networks also arises in other settings. For instance, the data structure typical of scientific networks is commonly found in studies of patent citation networks connecting teams of inventors citing related patents \citep{acemoglu2016innovation}. Similarly, artistic productions where multiple artists are involved in collaborative productions make explicit or implicit references to other productions or artists \citep{lena2004meaning, schuster2019sampling,burgdorf2024communities}. Many of the statistical effects that we have derived, implemented and tested in this paper in the context of scientific networks seem to arise naturally from repeated collaboration in artistic productions \citep{uzzi2005collaboration, faulkner1983music}, and music more specifically \citep{gleiser2003community}. These different collaborative contexts tend to create communities with permeable boundaries whose members reassemble over time in different combinations around different productions. We offer the conjecture that collaborative production of ideas tends to produce networks with very similar structural features almost regardless of contextual details \citep{saavedra2009simple}. 

Avenues for future research include both empirical, as well as methodological developments. Examples of empirical work that may benefit from the results reported in this paper include reproducibility studies of influential empirical findings in research on the science of science concerning, for example, the effect of citation atypicality on the impact of scientific papers \citep{uzzi2013atypical}, or the apparent decline of disruptive scientific innovations \citep{park2023papers}.  Using the more general models discussed in this paper, it may be possible to define new ways to normalize quantitative measures in citation networks and test the robustness of existing results across normalization strategies. Other possible empirical studies employing models proposed in this work could test theories of social selection (e.\,g., scientists start collaboration due to common research interest), social influence (e.\,g., scientists adopt research orientation of their current or past collaborators), formation of scientific teams, or determinants of scientific productivity, innovation, and impact.

Possibilities for future progress on the modeling front include developing a more exhaustive list of covariates for mixed two-mode networks, techniques to find optimal scaling or normalization of network attributes and covariates, and techniques to find optimal half-life periods for the decaying influence of past events. Model estimation could be made more robust, for instance, through Markov chain simulation to more efficiently explore the large space of (plausible) controls. Unobserved heterogeneity of authors and/or papers may be accounted for by adapting ideas from multilevel REM \citep[e.\,g.,][]{uzaheta2023random} or latent-space REM \citep[e.\,g.,][]{mulder2021latent} to RHEM proposed in this paper.


\section*{Acknowledgments}
Financial support for this work was provided by: the Deutsche Forschungsgemeinschaft (DFG) -- through grant number 321869138 to J.L.;  The Executive Agency for Financing Higher Education, Research, through Development and Innovation Funding UEFISCDI grant, code PN-III-P4-ID-PCE-2020-2828) to M.-G.H., and the Schweizerischer Nationalfonds (SNF) through grant number 100013\_192549 to A.L.




\appendix

\section{Separation into coauthoring and citation model}
\label{app:separate_models_details}

\subsection{The coauthoring model}
\label{sec:author_model}

We define the coauthoring model by focusing on the differences to the joint model defined above. We adapt the notation by using the superscript $^{(a)}$ whenever necessary (where 'a' stands for 'authors'). The coauthoring model is for counting processes on $\mathbb{R}_+\times\mathcal{P}(\mathcal{I})$, where the intensity on a set of authors $I\in\mathcal{P}(\mathcal{I})$ is specified as
\begin{equation}
  \label{eq:lambda_a}
\lambda^{(a)}_t(I)=\overline{\lambda}^{(a)}_t(|I|)\exp\left\{\beta^{(a){\rm T}}_0x^{(a)}_t(I)\right\}{\bf 1}\{I\subseteq\mathcal{I}_t\}\enspace.
\end{equation}
Note that, although covariates $x^{(a)}_t(I)$ are only functions of the set of authors $I$, this does not preclude that these covariates can consider \emph{past} citations. For example, $x^{(a)}_t(I)$ could count, for pairs of two different authors $i,i'\in I$, the number of papers that have been cited by both, $i$ and $i'$ in the past (this could indicate similarity of their research focus and could therefore increase the probability of a future collaboration); compare the precise definition of such a covariate in Section~\ref{sec:effects}. 
All covariates $x^{(a)}_t(I)$ of the coauthoring model can be used as covariates $x_t(I,J)$ in the joint model in a straightforward way by setting $x_t(I,J)=x^{(a)}_t(I)$, that is, by ignoring the second argument $J$.

For an observed sequence of publication events $(t_1,j_1,I_1,J_1),\dots,(t_n,j_n,I_n,J_n)$, the sampled partial likelihood $\tilde{L}^{(a)}_t(\beta)$ of the coauthoring model is very similar to the one of the joint model, given in (\ref{eq:likelihood_sampled}). The only remarkable difference is that the summation is over the sampled risk set $\tilde{\mathcal{R}}^{(a)}_{t_m}(\mathcal{I}_{t_m},I_m,q)$, sampled uniformly at random from
\begin{equation}
  \label{eq:sampled_risk_set_a}
  \left\{
  \mathcal{R}\subseteq
          {\mathcal{I}_{t}\choose |I|}
          \,;\;I\in\mathcal{R}\wedge|\mathcal{R}|=q+1
  \right\}\enspace,
\end{equation}
instead of being sampled from (\ref{eq:sampled_risk_set}).

\subsection{The citation model}
\label{sec:citation_model}

As before, we define the citation model by focusing on the differences to the joint model defined above, where we use the superscript $^{(c)}$ whenever necessary (where 'c' stands for 'citations'). Same as for the joint model, the citation model is for counting processes on $\mathbb{R}_+\times\mathcal{P}(\mathcal{I})\times\mathcal{P}(\mathcal{J})$. The first remarkable difference to the joint model is that the intensity on a pair $(I,J)\in\mathcal{P}(\mathcal{I})\times\mathcal{P}(\mathcal{J})$ absorbs the set of authors $I$ in the baseline intensity $\overline{\lambda}^{(c)}_t$:
\begin{equation}
  \label{eq:lambda_c}
\lambda^{(c)}_t(I,J)=\overline{\lambda}^{(c)}_t(I,|J|)\exp\left\{\beta_0^{(c){\rm T}}x^{(c)}_t(I,J)\right\}{\bf 1}\{I\subseteq\mathcal{I}_t\wedge J\subseteq\mathcal{J}_t\}\enspace.
\end{equation}
This difference implies that the hazard ratio $\exp\left\{\beta_0^{(c){\rm T}}x^{(c)}_t(I,J)\right\}$ explains by which factor a possible set of references $J$ is more or less likely than an alternative set of references $J'$, \emph{for a given publication by authors $I$.} In other words, the citation model conditions on the observed set of authors of a publication event and explains which references are likely to be cited by them.
The other implication is that any covariate $x^{(c)}_t(I,J)$ that is only a function of the authors $I$ (but ignores the references $J$) would lead to a non-identifiable parameter $\beta_0^{(c)}$ in the citation model, since its effect is absorbed in the baseline intensity. (From another perspective, the effect of such a covariate would cancel out in the resulting likelihood function.) For instance, a covariate $x^{(c)}_t(I,J)$ counting the number of previous papers of authors $i\in I$ would lead to a non-identifiable parameter, since it satisfies $x^{(c)}_t(I,J)=x^{(c)}_t(I,J')$ for all sets of references $J,J'$. On the other hand, a covariate $x^{(c)}_t(I,J)$ counting the number of those previous papers of authors $i\in I$ \emph{that cite papers in $J$} could be included in the citation model. All covariates of the citation model can also be included in the joint model.

The sampled partial likelihood $\tilde{L}^{(c)}_t(\beta)$ of the citation model is very similar to the one of the joint model, given in (\ref{eq:likelihood_sampled}). The only remarkable difference is that it uses in the summation the sampled risk set
$\tilde{\mathcal{R}}^{(c)}_t(I,\mathcal{J}_t,J,q)\subseteq\{I\}\times{\mathcal{J}_{t}\choose |J|}$ which is sampled uniformly at random from
\begin{equation}
  \label{eq:sampled_risk_set_c}
  \left\{
  \mathcal{R}\subseteq
          \{I\}\times{\mathcal{J}_{t}\choose |J|}
          \,;\;(I,J)\in\mathcal{R}\wedge|\mathcal{R}|=q+1
  \right\}\enspace,
\end{equation}
instead of being sampled from (\ref{eq:sampled_risk_set}). Note that the first component of the elements in $\tilde{\mathcal{R}}^{(c)}_t(I,\mathcal{J}_t,J,q)$ is fixed to $I$ and only the second component is random.

\section{Illustration of network effects}
\label{app:illustration}

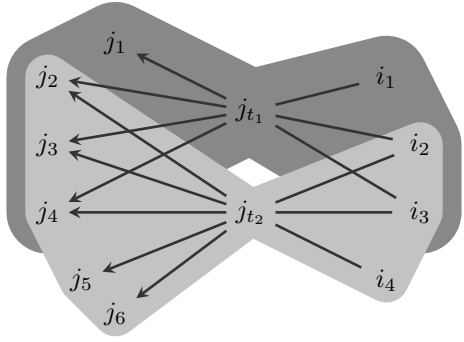
\begin{figure}
  \begin{center}
    \begin{tikzpicture}[scale=0.9,>=stealth]
      \tikzstyle{actor}=[circle,minimum size=1mm];
      \tikzstyle{work}=[rectangle,minimum size=1mm];
      \node[work] (w) at (0,0) {$j_{t_1}$};
      \node[work] (w1) at (-2.0,1.0) {$j_1$};
      \node[work] (w2) at (-3.0,0.5) {$j_2$};
      \node[work] (w3) at (-3.0,-0.5) {$j_3$};
      \node[work] (w4) at (-3.0,-1.5) {$j_4$};
      \node[work] (w5) at (-2.5,-2.5) {$j_5$};
      \node[work] (w6) at (-2.0,-3.0) {$j_6$};
      \node[work] (wt2) at (0,-1.5) {$j_{t_2}$};
      \node[actor] (a1) at (2,0.5) {$i_1$};
      \node[actor] (a2) at (2.5,-0.5) {$i_2$};
      \node[actor] (a3) at (2.5,-1.5) {$i_3$};
      \node[actor] (a4) at (2.0,-2.5) {$i_4$};
      
      \draw[color=darkgrey,line width=1pt] (w) to (a1);
      \draw[color=darkgrey,line width=1pt] (w) to (a2);
      \draw[color=darkgrey,line width=1pt] (w) to (a3);
      \draw[color=darkgrey,line width=1pt] (wt2) to (a2);
      \draw[color=darkgrey,line width=1pt] (wt2) to (a3);
      \draw[color=darkgrey,line width=1pt] (wt2) to (a4);

      \draw[color=darkgrey,line width=1pt,->] (w) to (w1);
      \draw[color=darkgrey,line width=1pt,->] (w) to (w2);
      \draw[color=darkgrey,line width=1pt,->] (w) to (w3);
      \draw[color=darkgrey,line width=1pt,->] (w) to (w4);

      \draw[color=darkgrey,line width=1pt,->] (wt2) to (w2);
      \draw[color=darkgrey,line width=1pt,->] (wt2) to (w3);
      \draw[color=darkgrey,line width=1pt,->] (wt2) to (w4);
      \draw[color=darkgrey,line width=1pt,->] (wt2) to (w5);
      \draw[color=darkgrey,line width=1pt,->] (wt2) to (w6);

      \begin{pgfonlayer}{background}
        \foreach \nodename in {w,w1,w2,w3,w4,w5,w6,wt2,a1,a2,a3,a4} {
          \coordinate (\nodename') at (\nodename);
        }
        \path[fill=grey1,draw=grey1,line width=1.1cm, line cap=round, line join=round] 
        (w') to (w1') to (w2') to (w3') to (w4')  to (w') to (a1') to (a2') to (a3') to (w') -- cycle;
        \path[fill=grey3,draw=grey3,line width=0.6cm, line cap=round, line join=round] 
        (wt2') to (w2') to (w3') to (w4') to (w5') to (w6')  to (wt2') to (a4') to (a3') to (a2') to (wt2') -- cycle;
      \end{pgfonlayer}
    \end{tikzpicture}
  \end{center}
  \caption{\label{fig:subrep} Example of two publication events $(t_1,j_{t_1},\{i_1,i_2,i_3\},\{j_1,j_2,j_3,j_4\})$ and $(t_2,j_{t_2},\{i_2,i_3,i_4\},\{j_2,j_3,j_4,j_5,j_6\})$ with $t_1<t_2$ (repeated from Fig.~\ref{fig:event}), used to illustrate several effects defined via subset repetition (see the text for a detailed explanation).}
\end{figure}

Below we illustrate the history-dependent covariates with small stylized examples represented in small graphics. For convenience, we repeat the definitions of the various effects that has already been provided in Section~\ref{sec:hyed_covs}. Several effect based on the generic subset-repetition covariate $subrep_t^{(k,\ell)}$ are illustrated via the example displayed in Fig.~\ref{fig:subrep}. In all examples we ignore the decay implied by the elapsed time of past events to provide simpler numeric results.

\paragraph{Prior papers and prior joint papers}
The average prior publication activity of a set of authors $I\subseteq\mathcal{I}$ is captured by the covariate
\[
\textup{prior papers}_t(I)=subrep_t^{(1,0)}(I,\emptyset)
\]
and is a measure of the past publication activity of authors in $I$. At $t_2$, the time of the second event in Fig.~\ref{fig:subrep}, it is 
\[
\textup{prior papers}_{t_2}(\{i_2,i_3,i_4\})=2/3\enspace,
\]
since two of the three authors have one prior publication and the third author has no prior publication.

Previous collaboration among pairs of authors is captured by 
\[
\textup{prior joint papers}_t(I)=subrep_t^{(2,0)}(I,\emptyset)\enspace,
\]
which averages the number of coauthored papers over all unordered pairs of authors in $I$. In the example of Fig.~\ref{fig:subrep} it is at the time of the second event
\[
\textup{prior joint papers}_{t_2}(\{i_2,i_3,i_4\})=1/3\enspace,
\]
since one of the three unordered pairs, namely $\{i_2,i_3\}$, has one prior joint publication and the other two pairs have none.

The covariate capturing the heterogeneity of past publication activity
\[
\textup{diff prior papers}_t(I)=\sum_{\{i',i''\}\in{I\choose 2}}
\frac{|cite^{(ap)}_t(\{i'\},\emptyset)-cite^{(ap)}_t(\{i''\},\emptyset)|}{{|I|\choose 2}}
\]
leads to the following value the example of Fig.~\ref{fig:subrep}
\[
\textup{diff prior papers}_{t_2}(\{i_2,i_3,i_4\})=2/3\enspace,
\]
since $i_2$ and $i_3$ have the identical prior publication activity (difference is zero) and the difference of past activity between $i_4$ and each of the other two is one.

\paragraph{Paper citation popularity and co-citation}
For a set of papers $J\subseteq\mathcal{J}$ the average number of past citations is captured by the covariate
\[
\textup{paper citation popularity}_t(J)=subrep_t^{(0,1)}(\emptyset,J)\enspace.
\]
In the example of Fig.~\ref{fig:subrep} it is
\[
\textup{paper citation popularity}_{t_2}(\{j_2,\dots,j_6\})=3/5\enspace,
\]
since three of the five papers $\{j_2,j_3,j_4,j_5,j_6\}$ have received one prior citation and the other two have received none.

Previous co-citation of pairs and triples of papers are captured by
\begin{eqnarray*}
\textup{paper-pair cocitation}_t(J)&=&subrep_t^{(0,2)}(\emptyset,J)\\
\textup{paper-triple cocitation}_t(J)&=&subrep_t^{(0,3)}(\emptyset,J)\enspace.
\end{eqnarray*}

In the example of Fig.~\ref{fig:subrep} we get
\[
\textup{paper-pair cocitation}_{t_2}(\{j_2,\dots,j_6\})=3/10
\]
since three unordered pairs, out of ${5 \choose 2}=10$, among the five papers $\{j_2,j_3,j_4,j_5,j_6\}$ have been cocited before and we get
\[
\textup{paper-triple cocitation}_{t_2}(\{j_2,\dots,j_6\})=1/10
\]
since one subset of size three, out of ${5 \choose 2}=10$, among the five papers $\{j_2,j_3,j_4,j_5,j_6\}$ have been cocited before. For comparison, at a later point in time $t_3>t_2$ we get $\textup{paper-triple cocitation}_{t_3}(\{j_1,j_4,j_5\})=0$, since the three papers have never been cocited by the same previous paper and we would get $\textup{paper-pair cocitation}_{t_3}(\{j_1,j_4,j_5\})=2/3$, since there are two pairwise co-citations, out of three possible ones.

\paragraph{Author citation repetition}
The hypothetical tendency of (groups of) authors to repeatedly cite the same papers can be tested via the following covariates.
\begin{eqnarray*}
\textup{author citation repetition}_t(I,J)&=&subrep_t^{(1,1)}(I,J)\\
\textup{author-pair citation repetition}_t(I,J)&=&subrep_t^{(2,1)}(I,J)\\
\textup{author-triple citation repetition}_t(I,J)&=&subrep_t^{(3,1)}(I,J)\enspace.
\end{eqnarray*}
In the example of Fig.~\ref{fig:subrep} we get at time $t_2$, writing $I_2=\{i_2,i_3,i_4\}$ and $J_2=\{j_2,j_3,j_4,j_5,j_6\}$
\begin{eqnarray*}
\textup{author citation repetition}_{t_2}(I_2,J_2)&=&\frac{2\cdot 3}{3\cdot 5}\\
\textup{author-pair citation repetition}_{t_2}(I_2,J_2)&=&\frac{1\cdot 3}{3\cdot 5}\\
\textup{author-triple citation repetition}_{t_2}(I_2,J_2)&=&\frac{0}{1\cdot 5}\\
\end{eqnarray*}
To see where these values come from we note that, for ``author citation repetition'' there are two authors ($i_2$ and $i_3$) who repeatedly cite three papers ($j_2$, $j_3$, and $j_4$) while publishing $j_{t_2}$, which yields the $2\cdot 3$ in the numerator. The number of possibilities of connecting an author from $I_2$ with a paper from $J_2$ is $3\cdot 5$, which yields the denominator. For ``author-pair citation repetition'' we observe that one unordered pair within $I_2$, namely $\{i_2,i_3\}$ repeatedly cites the three papers $j_2$, $j_3$, and $j_4$ and that the number of possibilities to connect an unordered pair of authors from $I_2$ with a paper from $J_2$ is $3\cdot 5$. For ``author-triple citation repetition'' we observed that no subset of three authors from $I_2$ repeatedly cites any paper from $J_2$, which explains the zero. For comparison, we further note that at a later point in time $t_3>t_2$ it would be 
\begin{eqnarray*}
\textup{author-triple citation repetition}_{t_3}(\{i_1,i_2,i_4\},J_2)&=&0\enspace.
\end{eqnarray*}
To see why this holds, we note that, although all three authors in $\{i_1,i_2,i_4\}$ have cited the papers $j_2$, $j_3$, and $j_4$ before $t_3$, they never did so in a joint publication, coauthored by all three of them.

\paragraph{Citing a paper and its references}
The tendency to adopt (some of) the references of a cited paper is captured by the following covariate measuring the past citation density within a set of papers $J\subseteq\mathcal{J}$:
\[
\textup{cite paper and its refs}_t(J)=\sum_{\{j',j''\}\in {J \choose 2}}
\frac{cite^{(pp)}_t(j',j'')+cite^{(pp)}_t(j'',j')}{|J|\cdot(|J|-1)}\enspace.
\]

\begin{figure}
  \begin{center}
    \begin{tikzpicture}[scale=0.9,>=stealth]
      \tikzstyle{actor}=[circle,minimum size=1mm];
      \tikzstyle{work}=[rectangle,minimum size=1mm];
      \node[work] (w) at (0,0) {$j_{t_1}$};
      \node[work] (w1) at (-2.0,1.0) {$j_1$};
      \node[work] (w2) at (-3.0,0.5) {$j_2$};
      \node[work] (w3) at (-3.0,-0.5) {$j_3$};
      \node[work] (w4) at (-3.0,-1.5) {$j_4$};
      \node[work] (w5) at (-2.5,-2.5) {$j_5$};
      \node[work] (w6) at (-2.0,-3.0) {$j_6$};
      \node[work] (wt2) at (1.5,-1.5) {$j_{t_2}$};
      
      \draw[color=darkgrey,line width=1pt,->] (w) to (w1);
      \draw[color=darkgrey,line width=1pt,->] (w) to (w2);
      \draw[color=darkgrey,line width=1pt,->] (w) to (w3);
      \draw[color=darkgrey,line width=1pt,->] (w) to (w4);

      \draw[color=darkgrey,line width=1pt,->] (wt2) to (w);
      \draw[color=darkgrey,line width=1pt,->] (wt2) to (w3);
      \draw[color=darkgrey,line width=1pt,->] (wt2) to (w4);
      \draw[color=darkgrey,line width=1pt,->] (wt2) to (w5);
      \draw[color=darkgrey,line width=1pt,->] (wt2) to (w6);

      \begin{pgfonlayer}{background}
        \foreach \nodename in {w,w1,w2,w3,w4,w5,w6,wt2} {
          \coordinate (\nodename') at (\nodename);
        }
        \path[fill=grey1,draw=grey1,line width=1.1cm, line cap=round, line join=round] 
        (w') to (w1') to (w2') to (w3') to (w4')  to (w') -- cycle;
        \path[fill=grey3,draw=grey3,line width=0.6cm, line cap=round, line join=round] 
        (wt2') to (w') to (w3') to (w4') to (w5') to (w6')  to (wt2') -- cycle;
      \end{pgfonlayer}
    \end{tikzpicture}
  \end{center}
  \caption{\label{fig:cite_paper_and_refs} Example of two publication events $(t_1,j_{t_1},I_1,\{j_1,j_2,j_3,j_4\})$ and $(t_2,j_{t_2},I_2,\{j_{t_1},j_3,j_4,j_5,j_6\})$ with $t_1<t_2$, illustrating the effect to cite a paper and part of its references. The author sets of these events do not matter for this covariate and are therefore left unspecified in the example.}
\end{figure}
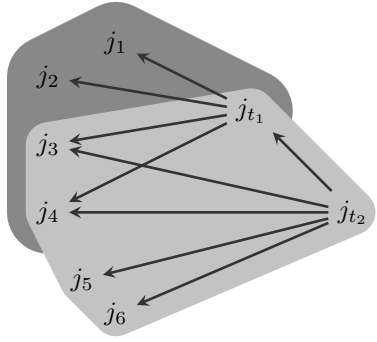

The covariate is illustrated in Fig.~\ref{fig:cite_paper_and_refs} where we get at time $t_2$
\[
\textup{cite paper and its refs}_{t_2}(\{j_{t_1},j_3,j_4,j_5,j_6\})=\frac{2}{5\cdot 4}\enspace,
\]
since there are $5\cdot 4$ possible past citation links within the set of five cited papers $\{j_{t_1},j_3,j_4,j_5,j_6\}$, two of which, namely $(j_{t_1},j_3)$ and $(j_{t_1},j_4)$, have been realized in a previous publication. (We may observe that, if publication times are given in a fine-grained scale, then reciprocal citations among papers are hardly possible, so that the maximum number of citation links should be divided by two. We argue that scaling covariates with a constant factor has no meaningful implications on the results, so that we can ignore this consideration.) 

\paragraph{Collaborate with citing author}
The tendency of scientists to coauthor papers with those who cited their previous work is captured by the covariate measuring the past citation density within a set of authors $I\subseteq\mathcal{I}$:
\[
\textup{collab w citing author}_t(I)=\sum_{\{i',i''\}\in {I \choose 2}}
\frac{cite^{(aa)}_t(i',i'')+cite^{(aa)}_t(i'',i')}{|I|\cdot(|I|-1)}\enspace.
\]

\begin{figure}
  \begin{center}
    \begin{tikzpicture}[scale=0.9,>=stealth]
      \tikzstyle{actor}=[circle,minimum size=1mm];
      \tikzstyle{work}=[rectangle,minimum size=1mm];
      \node[work] (w) at (0,1) {$j_{t_1}$};
      \node[work] (w1) at (-2.0,1.0) {$j_1$};
      \node[work] (w2) at (-3.0,0.5) {$j_2$};
      \node[work] (w3) at (5.5,-0.5) {$j_3$};
      \node[work] (w4) at (5.5,-1.5) {$j_4$};
      \node[work] (w5) at (-1.5,-3.0) {$j_5$};
      \node[work] (wt3) at (4,-1) {$j_{t_3}$};
      \node[work] (wt2) at (0,-3) {$j_{t_2}$};
      \node[actor] (a1) at (2,0.5) {$i_1$};
      \node[actor] (a2) at (2.5,-0.5) {$i_2$};
      \node[actor] (a3) at (2.5,-1.5) {$i_3$};
      \node[actor] (a4) at (2.0,-2.5) {$i_4$};
      
      \draw[color=darkgrey,line width=1pt] (w) to (a1);
      \draw[color=darkgrey,line width=1pt] (w) to (a2);

      \draw[color=darkgrey,line width=1pt] (wt3) to (a2);
      \draw[color=darkgrey,line width=1pt] (wt3) to (a3);

      \draw[color=darkgrey,line width=1pt] (wt2) to (a3);
      \draw[color=darkgrey,line width=1pt] (wt2) to (a4);

      \draw[color=darkgrey,line width=1pt,->] (w) to (w1);
      \draw[color=darkgrey,line width=1pt,->] (w) to (w2);

      \draw[color=darkgrey,line width=1pt,->] (wt3) to (w3);
      \draw[color=darkgrey,line width=1pt,->] (wt3) to (w4);

      \draw[color=darkgrey,line width=1pt,->] (wt2) to (w);
      \draw[color=darkgrey,line width=1pt,->] (wt2) to (w5);

      \begin{pgfonlayer}{background}
        \foreach \nodename in {w,w1,w2,w3,w4,w5,wt2,wt3,a1,a2,a3,a4} {
          \coordinate (\nodename') at (\nodename);
        }
        \path[fill=grey1,draw=grey1,line width=1.1cm, line cap=round, line join=round] 
        (w') to (w1') to (w2') to (w') to (a1') to (a2') to (w') -- cycle;
        \path[fill=grey5,draw=grey5,line width=0.4cm, line cap=round, line join=round] 
        (wt3') to (w3') to (w4') to (wt3') to (a3') to (a2') to (wt3') -- cycle;
        \path[fill=grey3,draw=grey3,line width=0.75cm, line cap=round, line join=round] 
        (wt2') to (w') to (w5') to (wt2') to (a4') to (a3') to (wt2') -- cycle;
      \end{pgfonlayer}
    \end{tikzpicture}
  \end{center}
  \caption{\label{fig:collab_w_citing_author} Example of three publication events $(t_1,j_{t_1},\{i_1,i_2\},\{j_1,j_2\})$, $(t_2,j_{t_2},\{i_3,i_4\},\{j_{t_1},j_5\})$, and $(t_3,j_{t_3},\{i_2,i_3\},\{j_3,j_4\})$, with $t_1<t_2<t_3$, illustrating the effect to collaborate with those authors who cited the own previous work.}
\end{figure}
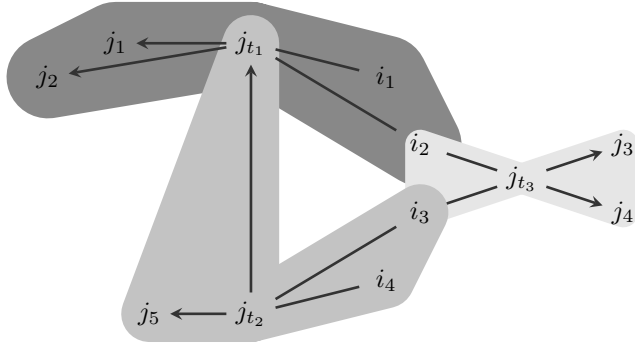

In the example from Fig.~\ref{fig:collab_w_citing_author} we get at $t_3$
\[
\textup{collab w citing author}_{t_3}(\{i_2,i_3\},\{j_3,j_4\})=1/2\enspace.
\]
To see why this is the case, recall the definition of the author-to-author citation attribute:
\[
cite^{(aa)}_t(i,i')=\sum_{t_m<t}w(t-t_m)\cdot{\bf 1}(i\in I_m\wedge \exists t_{m'}\colon j_{m'}\in J_m\wedge i'\in I_{m'})\enspace.
\]
and note that while publishing $j_{t_2}$, author $i_3$ cites the paper $j_{t_1}$ that has been published by $i_2$. Thus, after the publication of $j_{t_2}$ it is $cite^{(aa)}_t(i_3,i_2)=1$. This gives an author-to-author citation density of $1/2$ among $\{i_2,i_3\}$ at $t_3$. This example also illustrates the usefulness of the network attributes defined in Section~\ref{sec:attributes}. Judging from the raw data in Fig.~\ref{fig:collab_w_citing_author}, the effect ``collaborate with citing author'' is related with the closure of a cycle of length five, which is complex to understand and time-consuming to compute. With the network attribute $cite^{(aa)}$, which itself is intuitive to understand and can be explicitly stored while computing the covariates, the same effect becomes a mere computation of weighted density in a given set of authors.

We do not illustrate the effects ``author citation popularity'' or the heterogeneity in popularity with graphical examples, since we believe that these covariates are simple summary statistics of a node-level attribute and therefore easy to understand.

\paragraph{Author self citation}
Capturing the tendency of authors to cite their own past papers, we define the covariate \emph{self citation} measuring the density of the two-mode subgraph connecting a set of authors $I\subseteq\mathcal{I}$ and a set of papers $J\subseteq\mathcal{J}$ with respect to the $author$ relation
\[
\textup{self citation}_t(I,J)=\sum_{i'\in I,\;j'\in J}
\frac{author_t(i',j')}{|I|\cdot |J|}\enspace.
\]

\begin{figure}
  \begin{center}
    \begin{tikzpicture}[scale=0.9,>=stealth]
      \tikzstyle{actor}=[circle,minimum size=1mm];
      \tikzstyle{work}=[rectangle,minimum size=1mm];
      \node[work] (w) at (0,0) {$j_{t_1}$};
      \node[work] (w1) at (-2.0,1.0) {$j_1$};
      \node[work] (w2) at (-3.0,0.5) {$j_2$};
      \node[work] (w3) at (-3.0,-0.5) {$j_3$};
      \node[work] (w4) at (-3.0,-1.5) {$j_4$};
      \node[work] (w5) at (-2.5,-2.5) {$j_5$};
      \node[work] (w6) at (-2.0,-3.0) {$j_6$};
      \node[work] (wt2) at (0,-2.5) {$j_{t_2}$};
      \node[actor] (a1) at (2,0.5) {$i_1$};
      \node[actor] (a2) at (2.5,-0.5) {$i_2$};
      \node[actor] (a3) at (2.5,-1.5) {$i_3$};
      \node[actor] (a4) at (2.0,-2.5) {$i_4$};
      
      \draw[color=darkgrey,line width=1pt] (w) to (a1);
      \draw[color=darkgrey,line width=1pt] (w) to (a2);
      \draw[color=darkgrey,line width=1pt] (w) to (a3);
      \draw[color=darkgrey,line width=1pt] (wt2) to (a2);
      \draw[color=darkgrey,line width=1pt] (wt2) to (a3);
      \draw[color=darkgrey,line width=1pt] (wt2) to (a4);

      \draw[color=darkgrey,line width=1pt,->] (w) to (w1);
      \draw[color=darkgrey,line width=1pt,->] (w) to (w2);
      \draw[color=darkgrey,line width=1pt,->] (w) to (w3);

      \draw[color=darkgrey,line width=1pt,->] (wt2) to (w);
      \draw[color=darkgrey,line width=1pt,->] (wt2) to (w4);
      \draw[color=darkgrey,line width=1pt,->] (wt2) to (w5);
      \draw[color=darkgrey,line width=1pt,->] (wt2) to (w6);

      \begin{pgfonlayer}{background}
        \foreach \nodename in {w,w1,w2,w3,w4,w5,w6,wt2,a1,a2,a3,a4} {
          \coordinate (\nodename') at (\nodename);
        }
        \path[fill=grey1,draw=grey1,line width=1.1cm, line cap=round, line join=round] 
        (w') to (w1') to (w2') to (w3')  to (w') to (a1') to (a2') to (a3') to (w') -- cycle;
        \path[fill=grey3,draw=grey3,line width=0.6cm, line cap=round, line join=round] 
        (wt2') to (w') to (w4') to (w5') to (w6')  to (wt2') to (a4') to (a3') to (a2') to (wt2') -- cycle;
      \end{pgfonlayer}
    \end{tikzpicture}
  \end{center}
  \caption{\label{fig:self_citation} Example of two publication events $(t_1,j_{t_1},\{i_1,i_2,i_3\},\{j_1,j_2,j_3\})$ and $(t_2,j_{t_2},\{i_2,i_3,i_4\},\{j_{t_1},j_4,j_5,j_6\})$, illustrating the ``self-citation'' covariate.}
\end{figure}
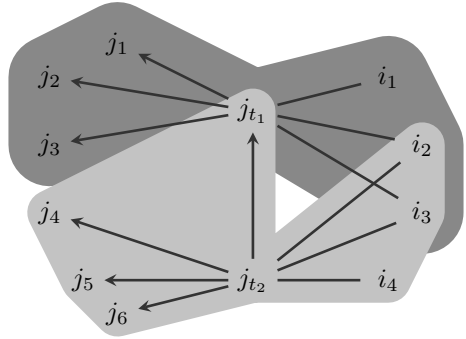

In the example of Fig.~\ref{fig:self_citation} we have at time $t_2$
\[
\textup{self citation}_{t_2}(\{i_2,i_3,i_4\},\{j_{t_1},j_4,j_5,j_6\})=\frac{2\cdot 1}{3\cdot 4}\enspace,
\]
since among the $3\cdot 4$ pairs linking an author to a cited paper, two link an author's own previous paper.

\paragraph{Coauthor closure} The covariate \emph{common coauthor} measures to what extent authors in $I\subseteq\mathcal{I}$ have coauthored papers with the same ``third'' authors. To abbreviate notation, we write $coauth_t(i',i'')=cite^{(ap)}_t(\{i',i''\},\emptyset)$.
\[
\textup{common coauthor}_t(I)=\sum_{\{i',i''\}\in{I\choose 2}\wedge i'''\neq i',i''}
\frac{\min[coauth_t(i',i'''),coauth_t(i'',i''')]}{{|I|\choose 2}}\enspace.
\]
In the summation above, the ``third'' author index $i'''$ runs over all authors $i'''\in\mathcal{I}$, different from the two focal authors $i'$ and $i''$.

\begin{figure}
  \begin{center}
    \begin{tikzpicture}[scale=1.0,>=stealth]
      \tikzstyle{actor}=[circle,minimum size=1mm];
      \tikzstyle{work}=[rectangle,minimum size=1mm];
      \node[work] (wt1) at (0,-0.3) {$j_{t_1}$};
      \node[work] (wt2) at (0,1.8) {$j_{t_2}$};
      \node[work] (wt3) at (3,0.8) {$j_{t_3}$};
      \node[actor] (a1) at (1.8,-1) {$i_1$};
      \node[actor] (a2) at (-2,0) {$i_2$};
      \node[actor] (a3) at (-2,1.5) {$i_3$};
      \node[actor] (a4) at (2.5,2) {$i_4$};
      \node[actor] (a5) at (1.8,3) {$i_5$};
      
      \draw[color=darkgrey,line width=1pt] (wt1) to (a1);
      \draw[color=darkgrey,line width=1pt] (wt1) to (a2);
      \draw[color=darkgrey,line width=1pt] (wt1) to (a3);

      \draw[color=darkgrey,line width=1pt] (wt2) to (a2);
      \draw[color=darkgrey,line width=1pt] (wt2) to (a3);
      \draw[color=darkgrey,line width=1pt] (wt2) to (a4);
      \draw[color=darkgrey,line width=1pt] (wt2) to (a5);

      \draw[color=darkgrey,line width=1pt] (wt3) to (a1);
      \draw[color=darkgrey,line width=1pt] (wt3) to (a4);

      \begin{pgfonlayer}{background}
        \foreach \nodename in {wt1,wt2,wt3,a1,a2,a3,a4,a5} {
          \coordinate (\nodename') at (\nodename);
        }
        \path[fill=grey1,draw=grey1,line width=1.1cm, line cap=round, line join=round] 
        (wt1') to (a2') to (a3') to (a1') to (wt1') -- cycle;
        \path[fill=grey3,draw=grey3,line width=0.8cm, line cap=round, line join=round] 
        (wt2') to (a3') to (a2') to (a4') to (a5') to (wt2') -- cycle;
        \path[fill=grey5,draw=grey5,line width=0.5cm, line cap=round, line join=round] 
        (wt3') to (a4') to (a1') to (wt3') -- cycle;
      \end{pgfonlayer}
    \end{tikzpicture}
  \end{center}
  \caption{\label{fig:common_coauthor} Example of three publication events $(t_1,j_{t_1},\{i_1,i_2,i_3\},J_1)$, $(t_2,j_{t_2},\{i_2,i_3,i_4,i_5\},J_2)$, and $(t_3,j_{t_3},\{i_1,i_4\},J_3)$, with $t_1<t_2<t_3$ illustrating the ``common coauthor'' covariate. References do not influence this covariate and are therefore not explicitly specified in this example.}
\end{figure}
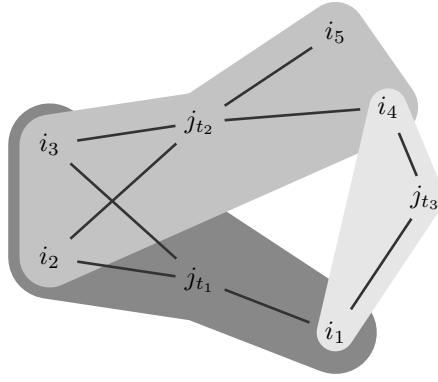

In the example of Fig.~\ref{fig:common_coauthor} we have at time $t_3$
\[
\textup{common coauthor}_{t_3}(\{i_1,i_4\})=2\enspace.
\]
The publication of $j_{t_1}$ establishes a coauthor link on $(i_1,i_2)$ and $(i_1,i_3)$ and the publication of $j_{t_2}$ establishes a coauthor link on $(i_4,i_2)$ and $(i_4,i_3)$. The two authors $i_2$ and $i_3$ take the role of the ``third author'' $i'''$ in the definition of the ``common coauthor'' covariate.

\paragraph{Citing common paper}
The covariate \emph{citing common paper} measures to what extent authors in $I\subseteq\mathcal{I}$ have cited the same papers in the past.
\[
\textup{cite common paper}_t(I)=\sum_{\{i',i''\}\in{I\choose 2}\wedge j\in\mathcal{J}}
\frac{\min[cite^{(ap)}_t(\{i'\},\{j\}),cite^{(ap)}_t(\{i''\},\{j\})]}{{|I|\choose 2}}\enspace.
\]

\begin{figure}
  \begin{center}
    \begin{tikzpicture}[scale=0.9,>=stealth]
      \tikzstyle{actor}=[circle,minimum size=1mm];
      \tikzstyle{work}=[rectangle,minimum size=1mm];
      \node[work] (w) at (0,1) {$j_{t_1}$};
      \node[work] (w1) at (-2.0,1.0) {$j_1$};
      \node[work] (w2) at (-3.0,0.5) {$j_2$};
      \node[work] (w3) at (5.5,-0.5) {$j_3$};
      \node[work] (w4) at (5.5,-1.5) {$j_4$};
      \node[work] (w5) at (-1.5,-3.0) {$j_5$};
      \node[work] (wt3) at (4,-1) {$j_{t_3}$};
      \node[work] (wt2) at (0,-3) {$j_{t_2}$};
      \node[actor] (a1) at (2,0.5) {$i_1$};
      \node[actor] (a2) at (2.5,-0.5) {$i_2$};
      \node[actor] (a3) at (2.5,-1.5) {$i_3$};
      \node[actor] (a4) at (2.0,-2.5) {$i_4$};
      
      \draw[color=darkgrey,line width=1pt] (w) to (a1);
      \draw[color=darkgrey,line width=1pt] (w) to (a2);

      \draw[color=darkgrey,line width=1pt] (wt3) to (a2);
      \draw[color=darkgrey,line width=1pt] (wt3) to (a3);

      \draw[color=darkgrey,line width=1pt] (wt2) to (a3);
      \draw[color=darkgrey,line width=1pt] (wt2) to (a4);

      \draw[color=darkgrey,line width=1pt,->] (w) to (w1);
      \draw[color=darkgrey,line width=1pt,->] (w) to (w2);

      \draw[color=darkgrey,line width=1pt,->] (wt3) to (w3);
      \draw[color=darkgrey,line width=1pt,->] (wt3) to (w4);

      \draw[color=darkgrey,line width=1pt,->] (wt2) to (w2);
      \draw[color=darkgrey,line width=1pt,->] (wt2) to (w5);

      \begin{pgfonlayer}{background}
        \foreach \nodename in {w,w1,w2,w3,w4,w5,wt2,wt3,a1,a2,a3,a4} {
          \coordinate (\nodename') at (\nodename);
        }
        \path[fill=grey1,draw=grey1,line width=1.1cm, line cap=round, line join=round] 
        (w') to (w1') to (w2') to (w') to (a1') to (a2') to (w') -- cycle;
        \path[fill=grey5,draw=grey5,line width=0.4cm, line cap=round, line join=round] 
        (wt3') to (w3') to (w4') to (wt3') to (a3') to (a2') to (wt3') -- cycle;
        \path[fill=grey3,draw=grey3,line width=0.75cm, line cap=round, line join=round] 
        (wt2') to (w2') to (w5') to (wt2') to (a4') to (a3') to (wt2') -- cycle;
      \end{pgfonlayer}
    \end{tikzpicture}
  \end{center}
  \caption{\label{fig:citing_common_paper} Example of three publication events $(t_1,j_{t_1},\{i_1,i_2\},\{j_1,j_2\})$, $(t_2,j_{t_2},\{i_2,i_3\},\{j_2,j_5\})$, and $(t_3,j_{t_3},\{i_2,i_3\},\{j_3,j_4\})$, with $t_1<t_2<t_3$ illustrating the ``citing common paper'' covariate. }
\end{figure}
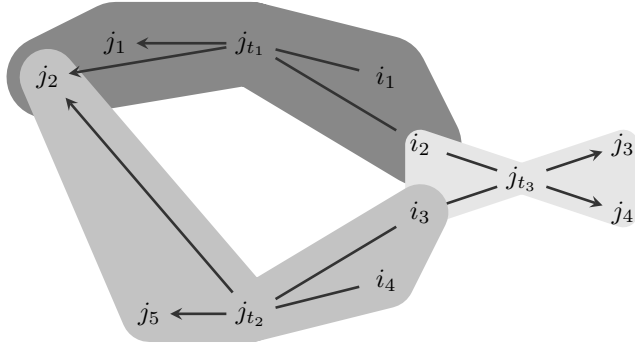

In the example of Fig.~\ref{fig:citing_common_paper} we have at $t_3$
\[
\textup{cite common paper}_{t_3}(\{i_2,i_3\})=1\enspace,
\]
since $i_2$ has cited paper $j_2$ (at $t_1$) and $i_3$ also has cited paper $j_2$ (at $t_2$).

\paragraph{Citing common author}
The covariate \emph{citing common author} measures to what extent authors in $I\subseteq\mathcal{I}$ have cited (papers published by) the same authors in the past.
\[
\textup{cite common auth}_t(I)=\sum_{\{i',i''\}\in{I\choose 2}\wedge i'''\neq i',i''}
\frac{\min[cite^{(aa)}_t(i',i'''),cite^{(aa)}_t(i'',i''')]}{{|I|\choose 2}}\enspace.
\]

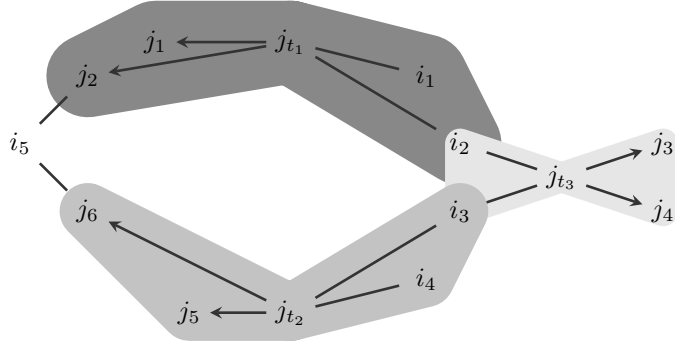
\begin{figure}
  \begin{center}
    \begin{tikzpicture}[scale=0.9,>=stealth]
      \tikzstyle{actor}=[circle,minimum size=1mm];
      \tikzstyle{work}=[rectangle,minimum size=1mm];
      \node[work] (w) at (0,1) {$j_{t_1}$};
      \node[work] (w1) at (-2.0,1.0) {$j_1$};
      \node[work] (w2) at (-3.0,0.5) {$j_2$};
      \node[work] (w6) at (-3.0,-1.5) {$j_6$};
      \node[work] (w3) at (5.5,-0.5) {$j_3$};
      \node[work] (w4) at (5.5,-1.5) {$j_4$};
      \node[work] (w5) at (-1.5,-3.0) {$j_5$};
      \node[work] (wt3) at (4,-1) {$j_{t_3}$};
      \node[work] (wt2) at (0,-3) {$j_{t_2}$};
      \node[actor] (a1) at (2,0.5) {$i_1$};
      \node[actor] (a2) at (2.5,-0.5) {$i_2$};
      \node[actor] (a3) at (2.5,-1.5) {$i_3$};
      \node[actor] (a4) at (2.0,-2.5) {$i_4$};
      \node[actor] (a5) at (-4.0,-0.5) {$i_5$};
      
      \draw[color=darkgrey,line width=1pt] (w) to (a1);
      \draw[color=darkgrey,line width=1pt] (w) to (a2);

      \draw[color=darkgrey,line width=1pt] (wt3) to (a2);
      \draw[color=darkgrey,line width=1pt] (wt3) to (a3);

      \draw[color=darkgrey,line width=1pt] (wt2) to (a3);
      \draw[color=darkgrey,line width=1pt] (wt2) to (a4);

      \draw[color=darkgrey,line width=1pt] (w2) to (a5);
      \draw[color=darkgrey,line width=1pt] (w6) to (a5);

      \draw[color=darkgrey,line width=1pt,->] (w) to (w1);
      \draw[color=darkgrey,line width=1pt,->] (w) to (w2);

      \draw[color=darkgrey,line width=1pt,->] (wt3) to (w3);
      \draw[color=darkgrey,line width=1pt,->] (wt3) to (w4);

      \draw[color=darkgrey,line width=1pt,->] (wt2) to (w6);
      \draw[color=darkgrey,line width=1pt,->] (wt2) to (w5);

      \begin{pgfonlayer}{background}
        \foreach \nodename in {w,w1,w2,w3,w4,w5,w6,wt2,wt3,a1,a2,a3,a4} {
          \coordinate (\nodename') at (\nodename);
        }
        \path[fill=grey1,draw=grey1,line width=1.1cm, line cap=round, line join=round] 
        (w') to (w1') to (w2') to (w') to (a1') to (a2') to (w') -- cycle;
        \path[fill=grey5,draw=grey5,line width=0.4cm, line cap=round, line join=round] 
        (wt3') to (w3') to (w4') to (wt3') to (a3') to (a2') to (wt3') -- cycle;
        \path[fill=grey3,draw=grey3,line width=0.75cm, line cap=round, line join=round] 
        (wt2') to (w6') to (w5') to (wt2') to (a4') to (a3') to (wt2') -- cycle;
      \end{pgfonlayer}
    \end{tikzpicture}
  \end{center}
  \caption{\label{fig:citing_common_author} Illustrating the ``citing common author'' covariate: example of three publication events $(t_1,j_{t_1},\{i_1,i_2\},\{j_1,j_2\})$, $(t_2,j_{t_2},\{i_3,i_4\},\{j_5,j_6\})$, and $(t_3,j_{t_3},\{i_2,i_3\},\{j_3,j_4\})$, with $t_1<t_2<t_3$. Moreover, papers $j_2$ and $j_6$ have both been authored by $i_5$ (in two publication events before $t_1$ that are not fully specified in this example).}
\end{figure}

In the example of Fig.~\ref{fig:citing_common_author} we have at $t_3$
\[
\textup{cite common auth}_{t_3}(\{i_2,i_3\})=1\enspace,
\]
since $i_2$ has cited the paper $j_2$ that has been authored by $i_4$ and $i_3$ has cited a paper, in this case $j_6$ that has also been authored by $i_4$.

\paragraph{Cited by common author}
Finally, the covariate \emph{cited by common author} reverses these relations and measures to what extent (papers published by) authors in $I\subseteq\mathcal{I}$ have been cited by the same authors in the past.
\[
\textup{cited by common auth}_t(I)=\sum_{\{i',i''\}\in{I\choose 2}\wedge i'''\neq i',i''}
\frac{\min[cite^{(aa)}_t(i''',i'),cite^{(aa)}_t(i''',i'')]}{{|I|\choose 2}}\enspace.
\]

\paragraph{Adopt citations of coauthors}
A tendency of authors to cite the papers that have been cited in the past by their past coauthors can be assessed by the following covariate
\[
\textup{adopt cite coauth}_t(I,J)=\sum_{i'\in I\wedge j'\in J \wedge i''\neq i'}
\frac{\min[coauth_t(i',i''),cite^{(ap)}_t(\{i''\},\{j'\})]}{|I| \cdot |J|}\enspace.
\]

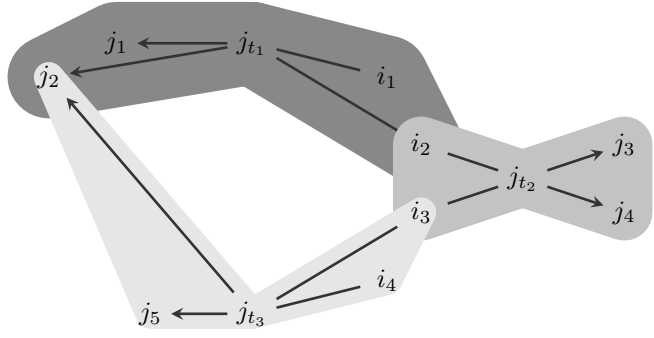
\begin{figure}
  \begin{center}
    \begin{tikzpicture}[scale=0.9,>=stealth]
      \tikzstyle{actor}=[circle,minimum size=1mm];
      \tikzstyle{work}=[rectangle,minimum size=1mm];
      \node[work] (w) at (0,1) {$j_{t_1}$};
      \node[work] (w1) at (-2.0,1.0) {$j_1$};
      \node[work] (w2) at (-3.0,0.5) {$j_2$};
      \node[work] (w3) at (5.5,-0.5) {$j_3$};
      \node[work] (w4) at (5.5,-1.5) {$j_4$};
      \node[work] (w5) at (-1.5,-3.0) {$j_5$};
      \node[work] (wt2) at (4,-1) {$j_{t_2}$};
      \node[work] (wt3) at (0,-3) {$j_{t_3}$};
      \node[actor] (a1) at (2,0.5) {$i_1$};
      \node[actor] (a2) at (2.5,-0.5) {$i_2$};
      \node[actor] (a3) at (2.5,-1.5) {$i_3$};
      \node[actor] (a4) at (2.0,-2.5) {$i_4$};
      
      \draw[color=darkgrey,line width=1pt] (w) to (a1);
      \draw[color=darkgrey,line width=1pt] (w) to (a2);

      \draw[color=darkgrey,line width=1pt] (wt2) to (a2);
      \draw[color=darkgrey,line width=1pt] (wt2) to (a3);

      \draw[color=darkgrey,line width=1pt] (wt3) to (a3);
      \draw[color=darkgrey,line width=1pt] (wt3) to (a4);

      \draw[color=darkgrey,line width=1pt,->] (w) to (w1);
      \draw[color=darkgrey,line width=1pt,->] (w) to (w2);

      \draw[color=darkgrey,line width=1pt,->] (wt2) to (w3);
      \draw[color=darkgrey,line width=1pt,->] (wt2) to (w4);

      \draw[color=darkgrey,line width=1pt,->] (wt3) to (w2);
      \draw[color=darkgrey,line width=1pt,->] (wt3) to (w5);

      \begin{pgfonlayer}{background}
        \foreach \nodename in {w,w1,w2,w3,w4,w5,wt2,wt3,a1,a2,a3,a4} {
          \coordinate (\nodename') at (\nodename);
        }
        \path[fill=grey1,draw=grey1,line width=1.1cm, line cap=round, line join=round] 
        (w') to (w1') to (w2') to (w') to (a1') to (a2') to (w') -- cycle;
        \path[fill=grey3,draw=grey3,line width=0.75cm, line cap=round, line join=round] 
        (wt2') to (w3') to (w4') to (wt2') to (a3') to (a2') to (wt2') -- cycle;
        \path[fill=grey5,draw=grey5,line width=0.4cm, line cap=round, line join=round] 
        (wt3') to (w2') to (w5') to (wt3') to (a4') to (a3') to (wt3') -- cycle;
      \end{pgfonlayer}
    \end{tikzpicture}
  \end{center}
  \caption{\label{fig:adopt_cite_of_coauthors} Illustrating the ``adopt citations of coauthors'' covariate: example of three publication events $(t_1,j_{t_1},\{i_1,i_2\},\{j_1,j_2\})$, $(t_2,j_{t_2},\{i_2,i_3\},\{j_3,j_4\})$, and $(t_3,j_{t_3},\{i_3,i_4\},\{j_2,j_5\})$, with $t_1<t_2<t_3$.}
\end{figure}

In the example of Fig.~\ref{fig:adopt_cite_of_coauthors} we have at $t_3$
\[
\textup{adopt cite coauth}_{t_3}(\{i_3,i_4\},\{j_2,j_5\})=1/4\enspace,
\]
since $i_2$ has cited the paper $j_2$, $i_2$ and $i_3$ become coauthors while publishing $j_{t_2}$ and then $i_3$ also cites $j_2$. The denominator is 4 since there are four author-paper pairs in the publication event $(t_3,j_{t_3},\{i_3,i_4\},\{j_2,j_5\})$.

\paragraph{Citing papers of coauthors}
Similarly, a tendency of authors to cite the papers that have been published by their past coauthors can be assessed by the following covariate
\[
\textup{cite paper of coauthor}_t(I,J)=\sum_{i'\in I\wedge j'\in J \wedge i''\neq i'}
\frac{\min[coauth_t(i',i''),author_t(i'',j')]}{|I| \cdot |J|}\enspace.
\]

\begin{figure}
  \begin{center}
    \begin{tikzpicture}[scale=0.9,>=stealth]
      \tikzstyle{actor}=[circle,minimum size=1mm];
      \tikzstyle{work}=[rectangle,minimum size=1mm];
      \node[work] (w) at (0,1) {$j_{t_1}$};
      \node[work] (w1) at (-2.0,1.0) {$j_1$};
      \node[work] (w2) at (-3.0,0.5) {$j_2$};
      \node[work] (w3) at (5.5,-0.5) {$j_3$};
      \node[work] (w4) at (5.5,-1.5) {$j_4$};
      \node[work] (w5) at (-1.5,-3.0) {$j_5$};
      \node[work] (wt2) at (4,-1) {$j_{t_2}$};
      \node[work] (wt3) at (0,-3) {$j_{t_3}$};
      \node[actor] (a1) at (2,0.5) {$i_1$};
      \node[actor] (a2) at (2.5,-0.5) {$i_2$};
      \node[actor] (a3) at (2.5,-1.5) {$i_3$};
      \node[actor] (a4) at (2.0,-2.5) {$i_4$};
      
      \draw[color=darkgrey,line width=1pt] (w) to (a1);
      \draw[color=darkgrey,line width=1pt] (w) to (a2);

      \draw[color=darkgrey,line width=1pt] (wt2) to (a2);
      \draw[color=darkgrey,line width=1pt] (wt2) to (a3);

      \draw[color=darkgrey,line width=1pt] (wt3) to (a3);
      \draw[color=darkgrey,line width=1pt] (wt3) to (a4);

      \draw[color=darkgrey,line width=1pt,->] (w) to (w1);
      \draw[color=darkgrey,line width=1pt,->] (w) to (w2);

      \draw[color=darkgrey,line width=1pt,->] (wt2) to (w3);
      \draw[color=darkgrey,line width=1pt,->] (wt2) to (w4);

      \draw[color=darkgrey,line width=1pt,->] (wt3) to (w);
      \draw[color=darkgrey,line width=1pt,->] (wt3) to (w5);

      \begin{pgfonlayer}{background}
        \foreach \nodename in {w,w1,w2,w3,w4,w5,wt2,wt3,a1,a2,a3,a4} {
          \coordinate (\nodename') at (\nodename);
        }
        \path[fill=grey1,draw=grey1,line width=1.1cm, line cap=round, line join=round] 
        (w') to (w1') to (w2') to (w') to (a1') to (a2') to (w') -- cycle;
        \path[fill=grey3,draw=grey3,line width=0.75cm, line cap=round, line join=round] 
        (wt2') to (w3') to (w4') to (wt2') to (a3') to (a2') to (wt2') -- cycle;
        \path[fill=grey5,draw=grey5,line width=0.4cm, line cap=round, line join=round] 
        (wt3') to (w') to (w5') to (wt3') to (a4') to (a3') to (wt3') -- cycle;
      \end{pgfonlayer}
    \end{tikzpicture}
  \end{center}
  \caption{\label{fig:cite_paper_of_coauthors} Illustrating the ``cite papers of coauthors'' covariate: example of three publication events $(t_1,j_{t_1},\{i_1,i_2\},\{j_1,j_2\})$, $(t_2,j_{t_2},\{i_2,i_3\},\{j_3,j_4\})$, and $(t_3,j_{t_3},\{i_3,i_4\},\{j_{t_1},j_5\})$, with $t_1<t_2<t_3$.}
\end{figure}
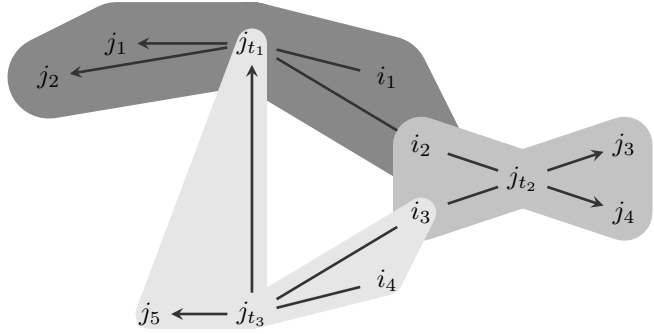

In the example of Fig.~\ref{fig:cite_paper_of_coauthors} we have at $t_3$
\[
\textup{cite paper of coauth}_{t_3}(\{i_3,i_4\},\{j_{t_1},j_5\})=1/4\enspace,
\]
since $i_2$ has authored the paper $j_{t_1}$, $i_2$ and $i_3$ become coauthors while publishing $j_{t_2}$ and then $i_3$ cites $j_{t_1}$. The denominator is 4 since there are four author-paper pairs in the publication event $(t_3,j_{t_3},\{i_3,i_4\},\{j_{t_1},j_5\})$.

\paragraph{Citation-repetition on the author level}
The covariate \emph{author-author citation repetition} can test the tendency of authors to repeatedly cite the papers of the same authors
\[
\textup{auth-auth cite repet}_t(I,J)=\sum_{i'\in I\wedge j'\in J \wedge i''\neq i'}
\frac{\min[cite^{(aa)}_t(i',i''),author_t(i'',j')]}{|I| \cdot |J|}\enspace.
\]

\begin{figure}
  \begin{center}
    \begin{tikzpicture}[scale=0.9,>=stealth]
      \tikzstyle{actor}=[circle,minimum size=1mm];
      \tikzstyle{work}=[rectangle,minimum size=1mm];
      \node[work] (w) at (0,1) {$j_{t_1}$};
      \node[work] (w1) at (-2.0,1.0) {$j_1$};
      \node[work] (w2) at (-3.0,0.5) {$j_2$};
      \node[work] (w3) at (5.5,-0.5) {$j_3$};
      \node[work] (w4) at (5.5,-1.5) {$j_4$};
      \node[work] (w5) at (-1.5,-3.0) {$j_5$};
      \node[work] (wt2) at (4,-1) {$j_{t_2}$};
      \node[work] (wt3) at (0,-1.5) {$j_{t_3}$};
      \node[work] (wt4) at (4,-2.5) {$j_{t_4}$};
      \node[actor] (a1) at (2,0.5) {$i_1$};
      \node[actor] (a2) at (2.5,-0.5) {$i_2$};
      \node[actor] (a3) at (2.5,-1.5) {$i_3$};
      \node[actor] (a4) at (2.5,-3.0) {$i_4$};
      
      \draw[color=darkgrey,line width=1pt] (w) to (a1);
      \draw[color=darkgrey,line width=1pt] (w) to (a2);

      \draw[color=darkgrey,line width=1pt] (wt2) to (a2);
      \draw[color=darkgrey,line width=1pt] (wt2) to (a3);

      \draw[color=darkgrey,line width=1pt] (wt3) to (a4);
      \draw[color=darkgrey,line width=1pt] (wt4) to (a4);

      \draw[color=darkgrey,line width=1pt,->] (w) to (w1);
      \draw[color=darkgrey,line width=1pt,->] (w) to (w2);

      \draw[color=darkgrey,line width=1pt,->] (wt2) to (w3);
      \draw[color=darkgrey,line width=1pt,->] (wt2) to (w4);

      \draw[color=darkgrey,line width=1pt,->] (wt3) to (w);
      \draw[color=darkgrey,line width=1pt,->] (wt3) to (w5);

      \draw[color=darkgrey,line width=1pt,->] (wt4) to (wt2);

      \begin{pgfonlayer}{background}
        \foreach \nodename in {w,w1,w2,w3,w4,w5,wt2,wt3,wt4,a1,a2,a3,a4} {
          \coordinate (\nodename') at (\nodename);
        }
        \path[fill=grey1,draw=grey1,line width=1.1cm, line cap=round, line join=round] 
        (w') to (w1') to (w2') to (w') to (a1') to (a2') to (w') -- cycle;
        \path[fill=grey2,draw=grey2,line width=0.8cm, line cap=round, line join=round] 
        (wt2') to (w3') to (w4') to (wt2') to (a3') to (a2') to (wt2') -- cycle;
        \path[fill=grey3,draw=grey3,line width=0.7cm, line cap=round, line join=round] 
        (wt3') to (w') to (w5') to (wt3') to (a4') to (wt3') -- cycle;
        \path[fill=grey5,draw=grey5,line width=0.4cm, line cap=round, line join=round] 
        (wt4') to (wt2')to (wt4') to (a4') to (wt4') -- cycle;
      \end{pgfonlayer}
    \end{tikzpicture}
  \end{center}
  \caption{\label{fig:auth-auth_cite_repet} Illustrating the ``author-author citation repetition'' covariate: }
\end{figure}
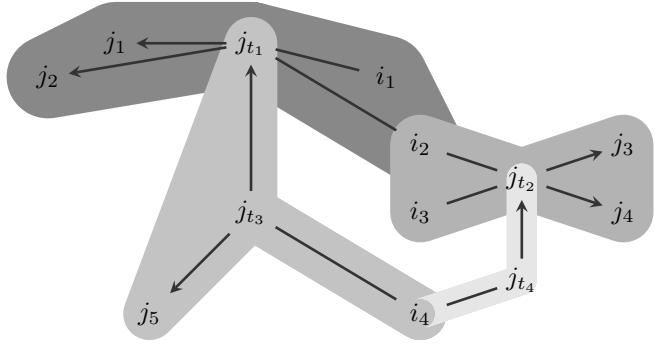

In the example of Fig.~\ref{fig:auth-auth_cite_repet} author $i_2$ publishes $j_{t_1}$ and $j_{t_2}$ in the first two events, author $i_4$ cites $i_2$'s paper $j_{t_1}$ while publishing $j_{t_3}$. Then, author $i_4$ repeatedly cites a paper of $i_2$ (in this case $j_{t_2}$) while publishing $j_{t_4}$. We have $\textup{auth-auth cite repetition}_{t_4}(\{i_4\},\{j_{t_2}\})=1$.

\paragraph{Citation-reciprocation on the author level}
Reversing a relation in the previously defined author-author citation repetition, we define a covariate \emph{author-author citation reciprocation} testing the tendency of authors to cite the papers of those other authors who have previously cited their papers
\[
\textup{auth-auth cite recipr}_t(I,J)=\sum_{i'\in I\wedge j'\in J \wedge i''\neq i'}
\frac{\min[cite^{(aa)}_t(i'',i'),author_t(i'',j')]}{|I| \cdot |J|}\enspace.
\]

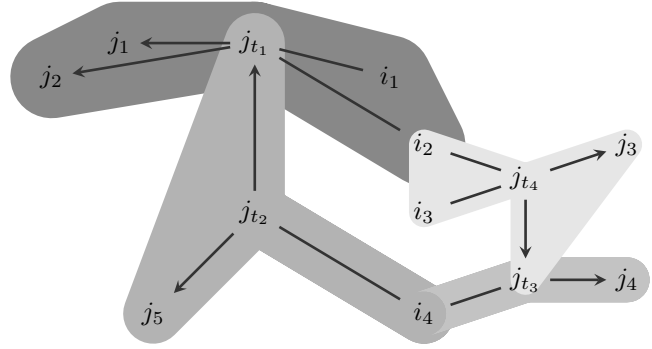
\begin{figure}
  \begin{center}
    \begin{tikzpicture}[scale=0.9,>=stealth]
      \tikzstyle{actor}=[circle,minimum size=1mm];
      \tikzstyle{work}=[rectangle,minimum size=1mm];
      \node[work] (w) at (0,1) {$j_{t_1}$};
      \node[work] (w1) at (-2.0,1.0) {$j_1$};
      \node[work] (w2) at (-3.0,0.5) {$j_2$};
      \node[work] (w3) at (5.5,-0.5) {$j_3$};
      \node[work] (w4) at (5.5,-2.5) {$j_4$};
      \node[work] (w5) at (-1.5,-3.0) {$j_5$};
      \node[work] (wt4) at (4,-1) {$j_{t_4}$};
      \node[work] (wt2) at (0,-1.5) {$j_{t_2}$};
      \node[work] (wt3) at (4,-2.5) {$j_{t_3}$};
      \node[actor] (a1) at (2,0.5) {$i_1$};
      \node[actor] (a2) at (2.5,-0.5) {$i_2$};
      \node[actor] (a3) at (2.5,-1.5) {$i_3$};
      \node[actor] (a4) at (2.5,-3.0) {$i_4$};
      
      \draw[color=darkgrey,line width=1pt] (w) to (a1);
      \draw[color=darkgrey,line width=1pt] (w) to (a2);

      \draw[color=darkgrey,line width=1pt] (wt4) to (a2);
      \draw[color=darkgrey,line width=1pt] (wt4) to (a3);

      \draw[color=darkgrey,line width=1pt] (wt2) to (a4);
      \draw[color=darkgrey,line width=1pt] (wt3) to (a4);

      \draw[color=darkgrey,line width=1pt,->] (w) to (w1);
      \draw[color=darkgrey,line width=1pt,->] (w) to (w2);

      \draw[color=darkgrey,line width=1pt,->] (wt4) to (w3);
      \draw[color=darkgrey,line width=1pt,->] (wt4) to (wt3);

      \draw[color=darkgrey,line width=1pt,->] (wt2) to (w);
      \draw[color=darkgrey,line width=1pt,->] (wt2) to (w5);

      \draw[color=darkgrey,line width=1pt,->] (wt3) to (w4);

      \begin{pgfonlayer}{background}
        \foreach \nodename in {w,w1,w2,w3,w4,w5,wt2,wt3,wt4,a1,a2,a3,a4} {
          \coordinate (\nodename') at (\nodename);
        }
        \path[fill=grey1,draw=grey1,line width=1.1cm, line cap=round, line join=round] 
        (w') to (w1') to (w2') to (w') to (a1') to (a2') to (w') -- cycle;
        \path[fill=grey2,draw=grey2,line width=0.8cm, line cap=round, line join=round] 
        (wt2') to (w') to (w5') to (wt2') to (a4') to (wt2') -- cycle;
        \path[fill=grey3,draw=grey3,line width=0.6cm, line cap=round, line join=round] 
        (wt3') to (w4')to (wt3') to (a4') to (wt3') -- cycle;
        \path[fill=grey5,draw=grey5,line width=0.4cm, line cap=round, line join=round] 
        (wt4') to (w3') to (wt3') to (wt4') to (a3') to (a2') to (wt4') -- cycle;
      \end{pgfonlayer}
    \end{tikzpicture}
  \end{center}
  \caption{\label{fig:auth-auth_cite_recip} Illustrating the ``author-author citation reciprocation'' covariate: }
\end{figure}

In the example of Fig.~\ref{fig:auth-auth_cite_recip} author $i_2$ publishes $j_{t_1}$. Author $i_4$ cites $i_2$'s paper $j_{t_1}$ while publishing $j_{t_2}$.
Author $i_4$ publishes $j_{t_3}$. Then, author $i_2$ cites the paper of $i_4$ while publishing $j_{t_4}$, reciprocating the previous author-author citation relation. We have $\textup{auth-auth cite reciprocation}_{t_4}(\{i_2,i_3\},\{j_{t_3},j_3\})=1/4$.

\end{document}